\documentclass[sigconf]{acmart}

\usepackage{amsmath}
\usepackage{amsthm}
\usepackage{color}
\usepackage{listings}
\usepackage{xspace}
\usepackage{enumitem}
\usepackage{hhline}
\usepackage{hyperref}
\usepackage{multirow}
\usepackage{tabularx}
\usepackage{epsfig}
\usepackage{algorithm}
\usepackage{color}
\usepackage{framed}
\usepackage{balance}
\usepackage{diagbox}
\usepackage{algpseudocode}
\usepackage[T1]{fontenc}
\usepackage[utf8]{inputenc}

\definecolor{codegreen}{rgb}{0,0.6,0}
\definecolor{codegray}{rgb}{0.5,0.5,0.5}
\definecolor{codepurple}{rgb}{0.58,0,0.82}
\definecolor{backcolour}{rgb}{0.95,0.95,0.92}
\lstdefinestyle{mystyle}{
    backgroundcolor=\color{backcolour},
    commentstyle=\color{codegreen},
    keywordstyle=\color{magenta},
    numberstyle=\tiny\color{codegray},
    stringstyle=\color{codepurple},
    basicstyle=\footnotesize\ttfamily,
    breakatwhitespace=false,
    breaklines=true,
    captionpos=b,
    keepspaces=true,
    numbers=left,
    numbersep=5pt,
    showspaces=false,
    showstringspaces=false,
    showtabs=false,
    tabsize=2,
    columns=fixed
}
\newcommand{\tool}{\textsc{NeuEx}\xspace}

\renewcommand{\paragraph}{\vspace{3pt}\noindent\textbf}

\newcommand{\RNum}[1]{\uppercase\expandafter{\romannumeral #1\relax}}
\algnewcommand{\algorithmicgoto}{\textbf{go to}}%
\algnewcommand{\Goto}[1]{\State \algorithmicgoto~\ref{#1}}
\algnewcommand{\SGoto}{\State \algorithmicgoto\xspace\xspace}%
\algnewcommand{\Label}{\State\unskip}

\lstdefinelanguage{JavaScript}{
 keywords={typeof, static, new, begin, end, struct, char, void, unsigned, long, const, privilege_enclave, int, true, false, catch, function, return, null, catch, switch, var, if, in, while, do, else, case, break},
 keywordstyle=\color{blue},
 ndkeywords={class, export, boolean, throw, implements, import, this},
 ndkeywordstyle=\color{darkgray}\bfseries,
 identifierstyle=\color{black},
 sensitive=false,
 comment=[l]{//},
 morecomment=[s]{/*}{*/},
 commentstyle=\color{purple}\ttfamily\bfseries,
 stringstyle=\color{red}\ttfamily,
 morestring=[b]',
 morestring=[b]"
}

\lstdefinestyle{JavaScript}{
   language={JavaScript}, 
   moredelim=**[is][\btHL]{`}{`},
   moredelim=**[is][{\btHL[fill=green!30,]}]{@}{@},
}

\lstdefinestyle{nonumbers}
{numbers=none}

\begin{document}
\title{Neuro-Symbolic Execution: \\The Feasibility of an Inductive Approach to Symbolic Execution}

\author{{\rm Shiqi Shen} \qquad {\rm Soundarya Ramesh} \qquad {\rm Shweta Shinde} \\
{\rm Abhik Roychoudhury} \qquad {\rm Prateek Saxena}}
\affiliation{National University of Singapore}
\affiliation{\textit{\{shiqi04, soundary, shweta24, abhik, prateeks\}@comp.nus.edu.sg}}

\begin{abstract}
Symbolic execution is a powerful technique for program analysis. However, it
has many limitations in practical applicability: the path explosion problem encumbers scalability, the need for language-specific implementation, the inability to handle complex dependencies, and the limited expressiveness of theories supported by underlying satisfiability checkers. Often, relationships between variables of
interest are not expressible directly as purely symbolic constraints. To this
end, we present a new approach --- {\em neuro-symbolic execution} --- which
learns an approximation of the relationship as a neural net. It features a
constraint solver that can solve mixed constraints, involving both symbolic
expressions and neural network representation. To do so, we 
envision such constraint solving as procedure combining SMT solving and 
gradient-based optimization. We demonstrate the utility of neuro-symbolic execution 
in constructing exploits for buffer overflows. We report success on 13/14 programs
which have difficult constraints, known to require specialized extensions to symbolic
execution. In addition, our technique solves $100$\% of the given neuro-symbolic constraints in $73$ programs from standard verification and invariant synthesis benchmarks.
\end{abstract}

\maketitle

\section{Introduction}
\label{sec:introduction}

Symbolic execution is a code analysis technique which reasons about sets of input values that drive the program to a specified state~\cite{king1976symbolic}. Certain inputs are marked as symbolic and the analysis gathers
symbolic constraints on these values, by analyzing the operations along a path of a program. Satisfying solutions to these constraints are concrete values that cause the program to execute the analyzed path leading to a particular state of interest. Manipulating these constraints allows one to reason about the reachability of different paths and states, thereby serving to guide search in the execution space efficiently.  Symbolic execution, especially its mixed-dynamic variant, has been widely used in computer security. Its prime application over the last decade has been in white-box fuzzing, with the goal of discovering software vulnerabilities~\cite{godefroid2012sage, godefroid2008automated, saxena2010symbolic}. More broadly, it
has been used for patching~\cite{perkins2009automatically, daniel2010test}, invariant discovery~\cite{gupta2009invgen}, and verification to prove the absence of vulnerabilities~\cite{jaffar2012tracer, coen2001using}. Off-the-shelf symbolic execution tools targeting languages such as C/C++~\cite{siegel2015civl}, JavaScript~\cite{li2014symjs,jalangi2}, Python~\cite{canini2012nice,PyExZ3}, and executable binary code~\cite{chipounov2011s2e} are available.

Symbolic analysis is a powerful technique; however, it has a number of limitations in practical applicability. First, symbolic analysis is mostly designed as a deductive procedure classically, requiring complete modeling of the target language (e.g., C vs. x64). A set of logical rules specific to the target language describe how to construct symbolic constraints for operations in that language~\cite{kudzu,jeon2012symdroid}. As new languages emerge, such symbolic analysis needs to be re-implemented for each language. More importantly, if a certain functionality of a program is unavailable for analysis --- either because it is implemented in a language different from the target language, or because it is accessible as a closed, proprietary service --- then, such functionality cannot be analyzed.

Second, the reasoning about symbolic constraints is limited to the expressiveness of theories supported by underlying satisfiability checkers (e.g.,  SAT / SMT solvers)~\cite{SurveySymExec}. Symbolic analysis typically uses quantifier-free and decidable theories in first-order logic, and satisfiability solvers have well-known limits~\cite{abraham2015building}. For instance, non-linear arithmetic over reals is not well supported in existing solvers, and string support is relatively new and still an area of active research~\cite{zheng2013z3,ganesh2011hampi}. When program functionality does not fall within the supported theories, analysis either precludes such functionality altogether, or encodes it abstractly using supported theories (e.g., arrays, bit-vectors, or uninterpreted functions).

Third, symbolic analyses often enumeratively analyze multiple paths in a program.  Complex control flows and looping structures are well-known to be missed by state-of-the-art implementations, which have attracted best-effort extensions to the basic technique and do not offer generality~\cite{LESE}. In particular, dynamic symbolic execution is known to suffer from scalability issues in long-running loops containing a large number of acyclic paths in the iterations, owing to loop unrolling and path explosion~\cite{cadar2013symbolic, xie2009fitness}.

\subsection{Neuro-Symbolic Execution}

In this paper, we aim to improve the expressiveness of symbolic execution to reason about parts of the code that are not expressible in the theories supported by the symbolic language (including its SMT theories), too complex, or simply unavailable in analyzable form. We present a technique called {\em neuro-symbolic execution}, which accumulates two types of constraints: standard {\em symbolic} constraints (derived deductively) and {\em neural} constraints (learned inductively). Neural constraints capture relations between program variables of code that are not expressible directly as purely symbolic constraints. The representation of these constraints is chosen to be a neural network (or neural net) in this work. The constraints including both symbolic and neural constraints are called {\em neuro-symbolic}.

Our procedure infers and manipulates neural constraints using only two generic interfaces, namely {\em learn} and {\em check satisfaction}. The first interface learns a neural network given concrete values of variables and an objective function to optimize. The second interface checks for satisfiability: given an output value for a neural network, finding whether an input evaluates to it. Both of these can be instantiated by many different procedures; we present a specific set of algorithms in this work for concreteness. We believe the general framework can be extended to other machine learning models which can implement such interfaces.

Our choice of representation via neural networks is motivated by two observations. First, neural nets can approximate or represent a large category of functions, as implied by the universal approximation theorem~\cite{funahashi1989approximate, hornik1991approximation}; and in practice, an explosion of empirical results are showing that they are learnable for many practical functions~\cite{andoni2014learning,godfrey2015continuum}. Although specialized training algorithms are continuously on the rise~\cite{qian1999momentum, kingma2014adam}, we expect that neural networks will prove effective in learning approximations to several useful functions we encounter in practice. Second, neural nets are a differentiable representation, often trained using optimization methods such as gradient descent~\cite{ruder2016overview}. This differentiability allows for efficient analytical techniques to check for satisfiability of neural constraints, and produce satisfying assignments of values to variables~\cite{goodfellow2014explaining,papernot2016limitations} --- analogous to the role of SMT solvers for purely symbolic constraints. One of the core technical contributions of this work is a procedure to solve {\em neuro-symbolic} constraints: checking satisfiability and finding assignments for variables involved in neural and symbolic constraints simultaneously, with good empirical accuracy on benchmarks tested.

Inductive synthesis of symbolic constraints usable in symbolic analyses has been attempted in prior work~\cite{DIG,DIG2,Daikon}. One notable difference is that our neural constraints are a form of {\em unstructured learning}, i.e.  they approximate a large class of functions and do {\em not} aim to print out constraints in a symbolic form amenable to SMT reasoning. Prior constraint synthesis works pre-determine a fixed template or structure of symbolic constraints --- for instance, octagonal inequalities~\cite{DIG}, low-degree polynomial equalities over integers~\cite{Daikon}, and so on. Each such template-based learning comes with a specialized learning procedure and either resorts to standard SMT solvers for solving constraints, or has hand-crafted procedures specialized to each template type. As a result, these techniques have found limited applicability in widely used symbolic execution analyses. 
As a side note, when the code being approximated does not fall within chosen template structure in prior works, they resolve to brute-force enumeration of templates to fit the samples.

\subsection{Applications \& Results}
Neuro-symbolic execution has the ability to reason about purely symbolic constraints, purely neural constraints, and mixed neuro-symbolic constraints. This approach has a number of possible future applications, including but not limited to: (a) analyzing protocol implementations without analyzable code~\cite{cui2007discoverer}; (b) analyzing code with complex dependency structures~\cite{xie2016proteus}; and (c) analyzing systems that embed neural networks directly as sub-components~\cite{bojarski2016end}.

To anchor our proposal, we focus on the core technique of neuro-symbolic execution through the lens of one application --- finding exploits for buffer overflows. In this setting, we show that neuro-symbolic execution can be used to synthesize neural constraints from parts of a program, to which the analysis only has black-box executable access. The program logic can have complex dependencies and control structure, and the technique does not need to know the operational semantics of the target language.
We show that for many real programs, our procedure can learn moderately accurate models, incorporate them with symbolic memory safety conditions, and solve them to uncover concrete exploits.

\paragraph{Tool.} We build a prototype tool (called \tool) to perform neuro-symbolic execution of C programs, where the analyst specifies which parts of the code it wants to treat as a black-box, and a memory unsafety condition (symbolic) which captures an exploit. \tool uses standard training algorithms to learn a neural net which approximates the black-box functionality and conjoins it with other symbolic constraints. Next, \tool employs a new procedure to solve the symbolic and neural constraints simultaneously, yielding satisfying assignments with high probability. The tool is constructive, in that it produces concrete values for free variables in the constraints, which can be tested as candidate exploits.

\paragraph{Results.} Our main empirical results are two-fold. First, we select a benchmark which has difficult constraints, known to require special-
ized extensions to symbolic execution. We show that \tool finds exploits for $13 / 14$ of programs in the benchmark. Our results are comparable to  binary-level symbolic execution tools~\cite{LESE} with little knowledge of the semantics of the target code and the specific language. The second empirical experiment analyzes two benchmarks used in prior works in invariant synthesis for verification and program synthesis~\cite{DIG,DIG2}. They comprise $73$ programs with $82$ loops and $259$ input variables in total. Given the neuro-symbolic constraints, \tool successfully solves 100\% neuro-symbolic constraints for these benchmarks.

\paragraph{Contributions.} We make the following contributions:
\begin{itemize}
\item {\em Neuro-Symbolic Constraints.} \tool represents the relationship between variables of code as a neural net without the knowledge of code semantics and language, and then conjoins it along with symbolic constraints.

\item {\em Neuro-Symbolic Constraint Solving.} \tool envisions constraint solving as a
search problem and encodes symbolic
constraints as an objective function for optimization along with neural net
to check their satisfiability.

\item {\em Evaluation.} \tool successfully constructs exploits for $13$ out of $14$ vulnerable programs, which is comparable to binary-level  symbolic execution~\cite{LESE}. In addition, \tool solves 100\% of given neuro-symbolic constraints over $73$ programs comprising of $259$ input variables in total.
\end{itemize} 
\section{Overview}\label{sec:problem}

\begin{figure}
\begin{lstlisting}[style=JavaScript, language=C, xleftmargin=0.3cm, captionpos=b]
void process_request(char * input){
  char URI[80], version[80], msgbuf[100];
  int ptr=0, uri_len=0, ver_len=0, i, j;
  if(strncmp(input, ''GET '', 4)!=0)
    fatal("Unsupported request!");
  while(input[ptr]!=' '){
    if(uri_len<80) URI[uri_len] = input[ptr];
    uri_len++; ptr++;
  }
  ptr++;
  while(input[ptr]!='\n'){
    if(ver_len<80) version[ver_len]=input[ptr];
    ver_len++; ptr++;
  }
  if(ver_len<8 || version[5]!='1')
    fatal('Unsupported protocol version');
  for(i=0,ptr=0; i<uri_len; i++,ptr++)
    msgbuf[ptr] = URI[i];
  msgbuf[ptr++] = ',';
  for(j=0;j<ver_len;j++,ptr++)
    msgbuf[ptr]=version[j]; // buffer overflow
  msgbuf[ptr++] = '\0'; // buffer overflow
  ...
}
\end{lstlisting}
\caption{A simplified example that parses the HTTP request and constructs a new message.}
\label{fig:motivating_example}
\end{figure}

Symbolic execution provides a tool that is useful in a variety of security-related applications. In this work, we focus on the challenges within symbolic execution and present a solution that is general for various kinds of programs. 

\subsection{Motivation and Challenges}

We outline a set of challenges posed to symbolic execution with the help of a real-world example from an HTTP server.

\paragraph{Motivating Example.}
Consider the simplified example of parsing the HTTP request shown in Figure~\ref{fig:motivating_example}. The code extracts the fields (e.g., uri and version) from the request and constructs the new message for further processing. Our goal is to check whether there exists any buffer overflow in this program. If so, we find the exploit that triggers the overflow. As shown in Figure~\ref{fig:motivating_example},  on Line 4-5, the function \texttt{process\_request} takes one input \texttt{input} and checks whether \texttt{input} starts with `\texttt{GET} '. On Line 6-14, it finds the \texttt{URI} and \texttt{version} from \texttt{input} by searching for the delimiter \texttt{` '} and \texttt{`$\backslash$n'} separately. Then, the function checks whether the program supports the request on Line 15-16 based on the \texttt{version}. Finally, it concatenates the \texttt{version} and \texttt{URI} with the delimiter \texttt{`,'} into a buffer \texttt{msgbuf} on Line 17-22. There exists a buffer overflow on Line 21-22, as the pointer \texttt{ptr} may exceed the boundary of \texttt{msgbuf}.

\paragraph{Challenge 1: Complex Dependency.}
To discover this buffer overflow via purely symbolic analysis, the technique has to reason about a complex dependency structure between the input and the variables of interest. Assume that the analyst has some knowledge of the input format, namely that the \texttt{input} has fields, \texttt{URI} and \texttt{version}, separated by \texttt{` '} and \texttt{`$\backslash$n'} and knows the allocated size of the \texttt{msgbuf} (which is \texttt{100}). By analyzing the program, the analyst knows the vulnerable condition of \texttt{msgbuf} is \texttt{ptr}$>99$, which leads to buffer overflows. Note that the path executed for reaching the vulnerability point on Line 21-22 involves updates to a number of variables (on Line 8 and 13) which do {\em not} have a direct dependency chain (rather a sequence of control dependencies) on the target variable \texttt{ptr}. Specifically, \texttt{uri\_len} and \texttt{ver\_len} are dependent on \texttt{input}, which in turn control \texttt{ptr} and the iterations of the vulnerable loop. Further, the relationship between \texttt{uri\_len}, \texttt{ver\_len}, and \texttt{ptr} involves reasoning over the conditional statements on Line 4 and 15, which may lead to the termination of function. Therefore, without specialized heuristics (e.g., loop-extension~\cite{LESE}), the state-of-the-art solvers resort to enumeration~\cite{klee}. For example, KLEE enumerates characters on \texttt{input} over \texttt{` '} and \texttt{`$\backslash$n'} until the input passes the checking on Line 15 and \texttt{ver\_len+uri\_len>98}.

The unavailability of source code is another challenge for capturing complex dependency between variables, especially when the functions are implemented as a remote call or a library call written in a different language. For example, a symbolic execution may abort for calls to native Java methods and unmanaged code in .NET, as the symbolic values flow outside the boundary of target code~\cite{anand2007type}. 
To handle this challenge, symbolic execution has to hard-code models for these unknown function calls, which requires considerable manual expertise. Even though symbolic execution tools often provide hand-crafted models for analyzing system calls, they do not precisely capture all the behaviors (e.g., the failure of system calls)~\cite{klee}. Thus, the constraints generated by purely symbolic execution cannot capture the real behavior of functions, which leads to the failure in vulnerability detection.

\paragraph{Challenge 2: Lack of Expressiveness.}
Additional challenges can arise in such analysis due to the complexity of the constraint and the lack of back-end theorem for solving the constraint.
As shown in Figure~\ref{fig:motivating_example}, the function is equivalent to a replacement based on regular expressions. It replaces the request of the form, \textquotedbl\texttt{GET\textvisiblespace}\textquotedbl \textit{URI} \textquotedbl\texttt{\textvisiblespace}\textquotedbl \textit{Version} \textquotedbl\texttt{$\backslash$n}\textquotedbl $\ast$, to the message of the form, \textit{URI} \textquotedbl\texttt{,}\textquotedbl \textit{Version} \textquotedbl\texttt{$\backslash$0}\textquotedbl on Line 4-22.\footnote{`$\ast$' matches as many characters as possible.} The complex relationship between the input and target buffer makes it infeasible for symbolic execution to capture it. Moreover, even if the regular expression is successfully extracted, the symbolic engine may not be able to solve it as the embedded SAT/SMT solver is not able to express certain theories (e.g., the string replacement and non-linear arithmetic). Although works have targeted these theories, current support for non-linear real and integer arithmetic is still in its infancy~\cite{abraham2015building}.

\subsection{Our Approach}
To address the above challenges, we propose a new approach with two main insights: (1) leveraging the high representation capability of neural nets to learn constraints when symbolic execution is infeasible to capture it; (2) encoding the symbolic constraints into neural constraint and leveraging the optimization algorithms to solve the neuro-symbolic constraints as a search problem.

\tool departs from the purist view that all variable dependencies and relations should be
{\em expressible precisely} in a symbolic form. Instead, \tool treats the entire code from Line 4-22 as a black-box, and inductively learn a neural network --- an approximate
representation of the logic mapping the variables of interest to target variables.
The constraint represented by the neural network is termed {\em neural constraint}. This neural constraint, say $N$, can represent relationships that may or may
not be representable as symbolic constraints. Instead, our approach creates a {\em neuro-symbolic} constraint, which includes both symbolic and neural constraints. Such neural constraint learning addresses the preceding first challenge as it learns the constraints from test data rather than source code.

Revisiting the example in Figure~\ref{fig:motivating_example}, the neuro-symbolic constraints capturing the vulnerability at the last control location on Line 22 are as follows.
\begin{equation}~\label{eq:eq1}
\texttt{uri\_length} = strlen(\textit{input\_uri})
\end{equation}
\begin{equation}
\texttt{ver\_length} = strlen(\textit{input\_version})
\end{equation}
\begin{equation}
\texttt{ptr} > 99
\end{equation}
\begin{equation}~\label{eq:eq4}
N: \{\texttt{uri\_length}, \texttt{ver\_length}\} \mapsto \{\texttt{ptr}\}
\end{equation}
where \texttt{uri\_length} is the length of uri field \textit{input\_uri} and \texttt{ver\_length} is the length of version field \textit{input\_version} in \texttt{input}. \footnote{\textit{input\_uri} and \textit{input\_version} are the content of the fields from input generated based on the knowledge of input, which is different from \texttt{URI} and \texttt{version} in Figure~\ref{fig:motivating_example}.} 
The first two constraints are symbolic constraints over the input fields, \texttt{uri\_length} and \texttt{ver\_length}.
The third symbolic constraint captures the vulnerable condition for \texttt{msgbuf}. The last constraint is a neural constraint capturing the relationship between the variable \texttt{uri\_length} and \texttt{ver\_length} and the variable \texttt{ptr} accessing the vulnerable buffer \texttt{msgbuf}.

\begin{table}[t]
\centering
\caption{The grammar of neuro-symbolic constraint language supported by \tool.}
\label{tbl:symbolic_grammar}
\begin{tabular}{cccl}
\hline
\begin{tabular}[c]{@{}c@{}}Nuero-Symbolic\\ Constraint\end{tabular} & 
\texttt{NS} & $\coloneqq$ & $N \land S$ \\ \hline
\begin{tabular}[c]{@{}c@{}}Neural\\ constraint\end{tabular} & 
\texttt{N} & $\coloneqq$ & $V_{I_n} \mapsto V_{O_n}$ \\ \hline
\begin{tabular}[c]{@{}c@{}}symbolic\\ constraint\end{tabular} & 
\texttt{S} & $\coloneqq$ & \texttt{e1} $\ominus$ \texttt{e2} | \texttt{e} \\ \hline
\multirow{2}{*}{Variable} & StrVar & $\coloneqq$ & ConstStr | StrVar$\circ$StrVar \\ 
 						  & NumVar & $\coloneqq$ & ConstNum | NumVar$\oslash$NumVar \\ \hline
\multirow{4}{*}{Expression} & \multirow{4}{*}{e} & \multirow{4}{*}{$\coloneqq$} 
       & contains(StrVar, StrVar) \\ 
 &  &  & strstr(StrVar, StrVar) $\otimes$ NumVar\\  
 &  &  & strlen(StrVar) $\otimes$ NumVar \\  
 &  &  & NumVar $\otimes$ NumVar \\ \hline
Logical     & $\ominus$ & $\coloneqq$ & $\lor$ | $\land$ \\ \hline
Conditional & $\otimes$ & $\coloneqq$ & == | $\neq$ | > | $\geq$ | < | $\leq$ \\ \hline
Arithmetic  & $\oslash$ & $\coloneqq$ & + | - | * | / \\ \hline
\end{tabular}
\end{table}

To the best of our knowledge, our approach is the first to train a neural net as a constraint and solve both symbolic constraint and neural constraint together. In our approach, we design an intermediate language, termed as neuro-symbolic constraint language. Table~\ref{tbl:symbolic_grammar} presents the syntax of the neuro-symbolic constraint language supported by \tool, which is expressive enough to model various constraints specified in many real applications such as string and arithmetic constraints.

Given the learned neuro-symbolic constraints, we seek the values of variables of interest that satisfy all the constraints within it. There exist multiple approaches to solve neuro-symbolic constraints. One naive way is to solve the neural and symbolic constraints separately. For example, consider the neuro-symbolic constraints in Equation~\ref{eq:eq1}-~\ref{eq:eq4}. We first solve the three symbolic constraints by SAT/SMT solvers and then discover a \textit{input\_uri} where \texttt{uri\_length}=10, a \textit{input\_version} where \texttt{ver\_length}=20 and a \texttt{ptr} whose value is 100.
Then, we feed the values of \texttt{uri\_length}, \texttt{ver\_length} and \texttt{ptr} to the neural constraint to check whether it satisfies the learned relationship. For the above case, the neural constraint produces the output such as 32 for the \texttt{ptr} when \texttt{uri\_length}=10 and \texttt{ver\_length}=20. Although this is a valid satisfiability result for the neural constraint, \texttt{ptr}=100 is not satisfiable for the current \textit{input\_uri} and \textit{input\_version}. This discrepancy arises because we solve these two types of constraints individually without considering the inter-dependency of variables within these constraints.
Alternatively, one could resort to enumeration over values of these three variables as a solution. However, it will require a lot of time for discovering the exploit.

This inspires our design of neuro-symbolic constraint solving. \tool's solving precedence is  
purely symbolic, purely neural and mixed constraints, in that order. 
Solving pure constraints is straightforward~\cite{z3, ruder2016overview}. 
The main technical novelty in our design is that \tool treats the mixed constraint solving as a search problem and utilizes the optimization algorithm to search for the satisfying solutions. To solve the mixed constraints simultaneously, \tool converts symbolic constraints to a loss function (or objective function) which is then used to guide the optimization of the loss function, thereby enabling  conjunction of symbolic and neural constraints.
\section{Design}
\label{sec:design}

\tool is the first tool to solve neuro-symbolic constraints. We first
explain the \tool setup and the building blocks we use in our approach.
Then, we present the core constraint solver of \tool along
with various optimization strategies.

\subsection{Overview}~\label{sec:design_overview}
\begin{figure}[t]
\centering
\epsfig{file=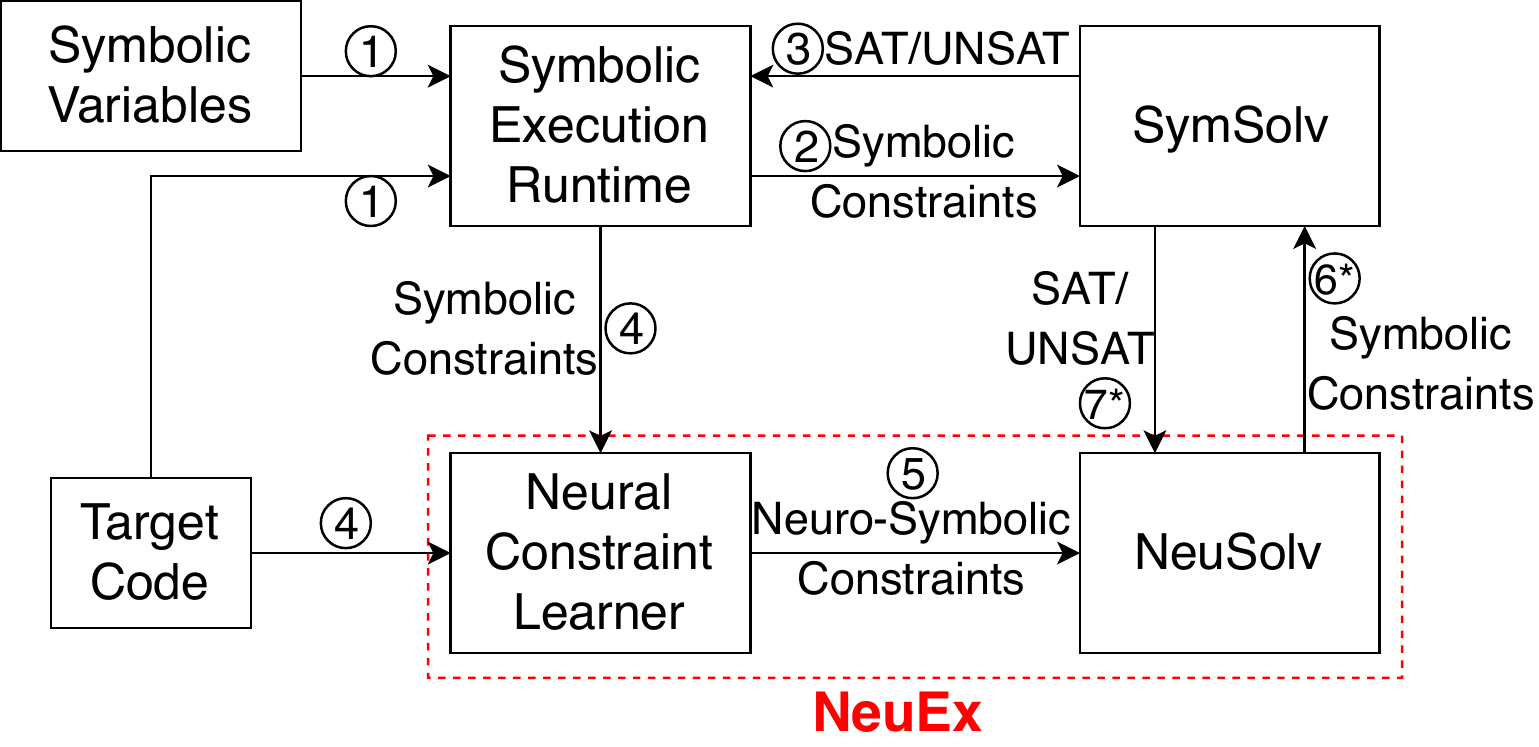, scale=0.55}
\caption{Workflow for \tool.  The circled number represents the order of operations. `*' represents the corresponding operations may be performed multiple times. SymSolv represents the symbolic constraint solver. NeuSolv represents the neuro-symbolic constraint solver.}
\label{fig:workflow}
\end{figure}

Symbolic execution is a generic technique to  automatically construct inputs
required to drive the program's execution  to a specific program point in the
code. To this end, a typical symbolic execution framework takes in a program
and a vulnerable condition for which we want to test the program. The analyst
using the framework also needs to mark the variables of interest as symbolic.
Typically, all the input variables are marked symbolic irrespective of its type. 
Further, environment variables, implicit inputs, user events, storage devices 
can also be marked as symbolic based on the use-case~\cite{exe, klee, kudzu}. 
Then the framework generates a set of test inputs to execute an execution path 
in the program. The analyst can aid this process by providing hints about input
grammar and so on, which they know beforehand. 

At each branch in the execution, the framework logs the symbolic constraints
collected so far as the path conditions required to reach this code point.
Specifically, a logical conjunction of all the symbolic constraints  gives us
the path constraints that have to be satisfied by the input to reach this code
point. Solving the symbolic path constraints gives us the concrete input value
which can  lead us to this execution point in the program, by invoking a constraint solver to produce concrete values. The framework may also negate the
symbolic constraints to explore other paths in the program or introduce a
feedback loop which uses the concrete input values returned by the constraint
solver as new inputs in order to increase path coverage.
Figure~\ref{fig:workflow} shows how \tool interacts with one such symbolic engine. 
It takes in symbolic constraint formulas in conjunctive normal form and returns
concrete values for each symbolic variable in the constraint formula if the
path is feasible; otherwise, it returns \texttt{UNSAT} which implies that the path
is infeasible. There are various solvers which support a wide range of
theories including but not limited to linear and non-linear arithmetic over integers,
booleans, bit-vectors, arrays and strings.

However, there does not exist any theory for solving neural constraints in
conjunction with symbolic constraints. A symbolic execution may need to solve
neural constraints for programs which invoke a neural network for parts of the
execution. For example,  if the web application uses a face recognition module
before granting access to a critical feature,  a traditional symbolic
framework will not be able to get past it.  Furthermore, symbolic execution is
well-known to fare badly for complex pieces of code involving
loops~\cite{LESE}.  Thus, whenever the symbolic engine's default constraint
solver is not able to find a solution and reaches its timeout, the framework can
pause its execution and automatically trigger an alternative mechanism. This
is where a  neural constraint solver comes into play. If the framework is
armed with a neural constraint solver such as \tool, it can model parts of the 
program a black-box and invoke the neural counterpart to solve the
constraints. Specifically, the framework can dispatch all the symbolic
constraints  it has collected so far along with the piece of code it wants to
treat as a black box. \tool in turn first adds all the symbolic constraints to
neuro-symbolic constraints and then queries its constraint solver to produce
concrete inputs or return \texttt{UNSAT}. In fact, any piece of code can be
modeled in terms of neural constraints to leverage \tool.  \tool is
generic in design as it can plug in any symbolic execution engine of choice.
It only requires the symbolic execution tool to provide two interfaces: one
for outputting the symbolic constraints and the other for querying the SAT/SMT
solvers as shown in Figure~\ref{fig:workflow}.
Table~\ref{tbl:symbolic_grammar} shows the grammar that \tool's constraint solver can reason about. For our example in
Figure~\ref{fig:motivating_example}, we want to infer the relations between
input HTTP request and variable index accessing the vulnerable buffer
\texttt{msgbuf}. So the symbolic framework will pass the following constraints
to \tool:
\begin{equation}~\label{eq:exampleconstraints}
\begin{split}
& \texttt{uri\_length} = strlen(\texttt{input\_uri})~\land \\
& \texttt{ver\_length} = strlen(\texttt{input\_version})~\land \\
& \texttt{ptr} > 99~\land\\
& N: \{\texttt{uri\_length}, \texttt{ver\_length}\} \mapsto \{\texttt{ptr}\}
\end{split}
\end{equation}

\subsection{Building Blocks} 
\label{subsec:building-blocks}

\tool's core engine solves the neuro-symbolic constraints such as in
Equation~\ref{eq:exampleconstraints} using its custom constraint solver
detailed in Section~\ref{sec:decision}.  It relies on two existing
techniques: SAT/SMT solver and gradient-based neural solver.
These solvers referred to as {\em SymSolv} and {\em NeuSolv} respectively form
the basic building blocks of \tool.

\paragraph{SymSolv.} 
\tool's symbolic constraint solver takes in first-order quantifier-free
formulas over multiple theories (e.g., empty theory, the theory of linear arithmetic and strings) and returns UNSAT or concrete values as output. It
internally employs Z3 Theorem Prover~\cite{z3} as an SMT solver to solve both
arithmetic and string symbolic constraints.

\paragraph{NeuSolv.}
For solving purely neural constraints, NeuSolv takes in the neural net and the
associated loss function to generate the expected values that
the output variables should have. \tool considers the neural constraint
solving as a search problem and uses a gradient-based search algorithm to
search for the satisfiable results. Gradient-based search algorithm searches for
the minimum of a given loss function $L(X)$ where $X$ is a n-dimensional vector~\cite{ruder2016overview}. The loss function can be any
differentiable function that monitors the error between the objective and
current predictions.  Consider the example in
Figure~\ref{fig:motivating_example}.  The objective of \tool is to check
whether the index \texttt{ptr} overruns the boundary of \texttt{msgbuf}.
Hence, the error is the distance between the value of \texttt{ptr} leading to
the buffer overflow and the value of \texttt{ptr} on Line 22 given by the
function \texttt{process\_request} with current input. By minimizing the
error, \tool can discover the input closest to the exploit. To minimize the
error, gradient-based search algorithm first starts with a random input $X_0$ which is the initial state of NeuSolv.
For every enumeration $i$, it computes the derivative $\nabla_{X_i}L(X_i)$
of $L(X_i)$ given the input $X_i$ and then update $X_i$ according to
$\nabla_{X_i}L(X_i)$. This is based on the observation that the derivative of
a function always points to a local nearest valley. The updated input
$X_{i+1}$ is defined as: 
\begin{equation}
X_{i+1} = X_i - \epsilon \nabla_{X_i}L(X_i)
\end{equation}
where
$\epsilon$ is the learning rate that controls how much is going to be updated.
Gradient-based search algorithm keeps updating the input until it reaches the
local minima. To avoid the non-termination case, we set the maximum number of enumerations to be $M_e$. If it exceed $M_e$, NeuSolv stops and returns current updated result.
Note that gradient-based search algorithm can only find the
local minima since it stops when the error increases. If the loss function is
a non-convex function with multiple local minima, the found local minima may
not be the global minima. Moreover, it may find different local minima when
the initial state is different. Thus, \tool executes the search
algorithm multiple times with different initial states in order to find the
global minima of $L(X)$. 

\begin{figure}[t]
\centering
\epsfig{file=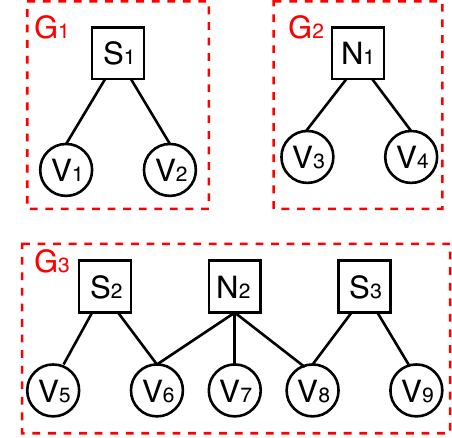, scale=1.1}
\caption{
\tool's DAG representation. S represents a symbolic constraint; N represents a
neural constraint; V represents a variable. The dotted rectangle represents
connected components of the DAG.
}
\label{fig:alg_example}
\end{figure}

\begin{algorithm}[H]
\caption{Algorithm for neuro-symbolic constraint solving. $S_p$ is purely symbolic constraints; $N_p$ is purely neural constraints; $S_m$ and $N_m$ are symbolic constraints and neural constraints in mixed components.}
\label{alg:const_solving}
\begin{algorithmic}[1]
\Function{NeuCL}{$S$, $N$} \Comment{$S$: Symbolic constraint list; $N$: Neural constraint list}
    \State ($S_p$, $N_p$, $S_m$, $N_m$) $\gets$ CheckDependency(N,S);
    \State (X, assign1) $\gets$ SymSolv($S_p$, $\emptyset$);
    \If{X == UNSAT}
        \State \Return (False, $\emptyset$);
    \EndIf
    \State cnt $\gets$ 0;
    \While{cnt<MAX\_TRIAL1} 
        \State (X, assign2) $\gets$ NeuSolv($N_p$);
        \If{X == SAT}
            \Goto{approach1}
        \EndIf
        \State cnt $\gets$ cnt+1;
    \EndWhile
    \SGoto{UNSAT}
    \State assign $\gets$ Union(assign1, assign2); \label{approach1}
    \State ConflictDB $\gets$ $\emptyset$; trial\_cnt $\gets$ 0;
    \While{trial\_cnt < MAX\_TRIAL2}
        \State ConflictConsts $\gets$ CreateConflictConsts(ConflictDB)
        \State (X, assign3) $\gets$ SymSolv($S_m$, ConflictConsts);
        \If{X == UNSAT}
            \SGoto{UNSAT}
        \EndIf
        \State NeuralConsts $\gets$ PartialAssign($N_m$, assign3); cnt $\gets$ 0;
        \While{cnt<MAX\_TRIAL1}
            \State (X, assign4) $\gets$ NeuSolv(NeuralConsts);
            \If{X == SAT}
                 \State assign2 $\gets$ Union(assign3, assign4); \SGoto{SAT}
            \EndIf
            \State cnt $\gets$ cnt+1;
        \EndWhile
        \State trial\_cnt $\gets$ trial\_cnt+1;
        \State ConflictDB $\gets$ ConflictDB $\cup$ assign3;
    \EndWhile
    \State trial\_cnt $\gets$ 0; F $\gets$ Encode($S_m$, $N_m$);
    \While{trial\_cnt<MAX\_TRIAL1}
        \State assign2 $\gets$ NeuSolv(F);
        \State X $\gets$ CheckSAT(assign2, $S_m$);
        \If{X == SAT}
            \SGoto{SAT}
        \EndIf
        \State trial\_cnt $\gets$ trial\_cnt+1;
    \EndWhile
    \Label \texttt{UNSAT:} 
        \State \Return (False, $\emptyset$);
    \Label \texttt{SAT:} 
        \State \Return (True, Union(assign, assign2))
\EndFunction

\end{algorithmic}
\end{algorithm}
\subsection{Constraint Solver}
~\label{sec:decision}

We propose a constraint solver to solve the neuro-symbolic
constraints with the help of SymSolv and NeuSolv. If the solver returns \texttt{SAT}, then the neuro-symbolic constraints guarantee to be satisfiable. It is not guaranteed to decide all satisfiability results with a timeout.
Algorithm~\ref{alg:const_solving} shows the algorithm for neuro-symbolic constraint solver. Interested readers can refer to Algorithm~\ref{alg:const_solving} for the precise algorithm.

\paragraph{DAG Generation.}
\tool takes the neuro-symbolic constraints and generates the directed acyclic
graph (DAG) between constraints and its variables. Each vertex of the DAG
represents a variable or constraint, and the edge shows that the variable is
involved in the constraint. For example, Figure~\ref{fig:alg_example} shows
the generated DAG for constraints $(V_1~op_1~V_2) \land (V_3~op_2~V_4)
\land (V_5~op_3~V_6) \land (V_6~op_3~V_7~op_4~V_8) \land (V_8~op_5~V_9)$ where $op_k$ can be any operator.

Next, \tool partitions the DAG into connected components by breadth-first
search~\cite{bundy1984breadth}. Consider the example shown in Figure~\ref{fig:alg_example}.
There are 5 constraints that are partitioned into three connected components,
$G_1$, $G_2$ and $G_3$.  \tool topologically sorts the components based on the
type of constraints to schedule the solving sequence. Specifically, it
clusters the components with only one kind of constraints as pure
components (e.g., $G_1$ and $G_2$) and the components including both constraints as
mixed components (e.g., $G3$). It further sub-categorizes pure
components into purely symbolic (e.g., $G_1$) and purely neural constraints (e.g.,
$G_2$).

\tool assigns solving precedence to be pure and mixed constraints. \tool
solves the mixed constraints in the end because the constraints have
different representation and hence are time-consuming to solve. Thus, in our
example, \tool first solves $S1 \land N1$ and then checks the
satisfiability of $S_2 \land N_2 \land S_3$. 


\paragraph{Pure Constraint Solving.} 
In pure constraints, we first apply SymSolv to solve purely symbolic constraints on Line 3 and then handle purely neural constraints using NeuSolv on Line 9. Note that the order of these two kinds of constraints does not affect the result. We solve purely symbolic constraints first because the 
SymSolv is fast, while the search algorithm for neural constraints requires numerous iteration and may not terminate. 
So, if the SymSolv reports \texttt{UNSAT} for purely symbolic
constraints, the whole neuro-symbolic constraints is \texttt{UNSAT}, as all the constraints are conjunctive.
Such an early \texttt{UNSAT} detection speeds up the satisfiability checking. If both solvers output \texttt{SAT}, \tool continues the process of solving the mixed
constraints.

\paragraph{Mixed Constraint Solving \RNum{1}.}
\tool obtains the symbolic constraints from mixed components (e.g.,
$S_2$ and $S_3$) by cutting the edges between the neural constraints and its
variables. Then, \tool invokes SymSolv to check their satisfiability on
Line 20. If the solver returns \texttt{UNSAT}, \tool goes to \texttt{UNSAT}
state; otherwise, \tool collects the concrete values of variables used in
these symbolic constraints. Then, \tool plugs these concrete values into
neural constraints on Line 24. For example, in Figure~\ref{fig:alg_example},
if the satisfiability result of $S_2 \land S_3$ is $<t_5, t_6, t_8, t_9>$ for
the variables $<V_5,V_6,V_8,V_9>$, \tool partially assigns $V_6$ and $V_8$ in
$N_2$ to be $t_6$ and $t_8$ separately. Now, we have the partially assigned neural constraint $N_2$\textquotesingle ~from $N_2$. All that remains is to
search for the value of $V_7$ satisfying $N_2$\textquotesingle.

To solve such a partially assigned neural constraint, \tool employs
NeuSolv on Line 26. If the NeuSolv
outputs \texttt{SAT}, \tool goes to \texttt{SAT} state. In \texttt{SAT} state,
\tool terminates and returns \texttt{SAT} with the combination of the
satisfiability results for all the constraints.   If the NeuSolv outputs \texttt{UNSAT}, \tool considers the satisfiability result of
symbolic constraints as a counterexample and derives a conflict clause on Line
19.  Specifically in our example, \tool creates a new conflict clause
$(V_5 \neq t_5) \lor (V_6 \neq t_6) \lor (V_8 \neq t_8) \lor (V_9 \neq t_9)$. Then
\tool adds this clause (Line 34) and queries the SymSolv with these new
symbolic constraints (Line 20). This method of adding conflict clauses is
similar to the backtracking in DPLL algorithm~\cite{davis1962machine}. Although the conflict clause learning approach used in \tool is simple, \tool is generic to adopt other advance strategies for constraint solving~\cite{silva1997grasp,moskewicz2001chaff, liang2016exponential}.

The above mixed constraint solving keeps executing the backtracking
procedure until it does not find any new counterexample. Consider the example
in Figure~\ref{fig:motivating_example}. \tool first finds the input whose
\texttt{uri\_length}=10, \texttt{ver\_length}=30, and \texttt{ptr}=100.
However, the result generated in this trial does not satisfy the neural
constraint. Then, \tool transforms this counterexample into a conflict clause
and goes to next trial to discover a new result. But this trial can be very expensive. For the example in Section~\ref{sec:problem},
mixed constraint solving takes more than $5000$ trials in the
worst case even after augmenting the constraints with additional information
that the value of \texttt{ptr} is $100$. To speed up mixed solving, \tool
chooses to limit the number of trials to a threshold value.

Specifically, if we do not have a SAT decision after mixed constraint solving
\RNum{1} within $k$ iterations\footnote{Users can adapt $k$ according to their applications.}, \tool applies an alternative strategy where we combine the symbolic constraints with neural constraints together. 
There exist two possible strategies: transforming neural constraints to symbolic constraints or the other way around. However, collapsing neural constraints to symbolic constraints incurs massive encoding clauses. For example, merely
encoding a small binarized neural network generates millions of variables and
millions of clauses~\cite{narodytska2017verifying}.
Thus, we transform the mixed constraints into purely neural constraints for solving them together.

\paragraph{Mixed Constraint Solving \RNum{2}.}
\tool collapses symbolic constraints to neural constraints by encoding the
symbolic constraints to a loss function on Line 36. This ensures the symbolic
and neural constraints are in the same form. For example, in
Figure~\ref{fig:alg_example}, \tool transforms the constraint $S_2$ and $S_3$
into a loss function of $N_2$.

Once the symbolic constraints are encoded into neural constraints, \tool
applies the NeuSolv to minimize the loss function on
Line 38. The main intuition behind this approach is to guide the search with
the help of encoded symbolic constraints. The loss function measures the
distance between current result and the satisfiability result of symbolic
constraints. The search algorithm gives us a candidate value for
satisfiability checking of neural constraints. However, the candidate value
generated by minimizing the distance may not always satisfy the symbolic
constraints since the search algorithm only tries to minimize the loss, rather
than exactly forces the satisfiability of symbolic constraints. To weed out such
cases, \tool checks the satisfiability for the symbolic constraints by
plugging in the candidate value and querying the SymSolv on Line 39.
If the result is \texttt{SAT}, \tool goes to \texttt{SAT} state. Otherwise,
\tool continues executing Approach \RNum{2} with a different initial state of
the search algorithm. For example, in Figure~\ref{fig:alg_example}, \tool
changes the initial value of $V_7$ for every iteration. Note that each
iteration in  Approach I has to execute sequentially because the addition of the
conflict clause forces serialization. As opposed to this, each trial
in Approach II is independent and thus embarrassingly parallelizable.

To avoid the non-termination case, \tool sets the maximum number of
trials for mixed constraint solving \RNum{2} to be $M_t$, which can be
configured independently of our constraint solver. Empirically, we notice
that the mixed constraint solving \RNum{2} is always able to find the
satisfiability result for complex constraints before hitting the threshold of
10.

\subsection{Encoding Mixed Constraints} 
\label{sec:encode}

\tool's mixed constraints take up most of the time during solving. 
We reduce this overhead by transforming them  to purely neural constraints. Specifically, \tool encodes the symbolic constraints $S(X)$ as a loss function $L(X)$
such as:
\begin{equation}~\label{eq:encode}
S(X) = S(min_X(L(X)))
\end{equation}
Next, \tool uses this loss function along with neural constraints and
applies NeuSolv to minimize the loss function of the entire mixed constraints. This encoding has two main advantages. First, it is straightforward to encode symbolic constraints into loss function. Second,
there exists gradient-based search algorithm for minimizing the loss function, which speeds up constraint solving in \tool.

\paragraph{Generic Encoding.}
As long as we have a loss function for the symbolic constraints, we can apply NeuSolv to solve the mixed constraints. In this paper, 
given the grammar of symbolic constraints shown in
Table~\ref{tbl:symbolic_grammar}, there exist six types of symbolic
constraints and two kinds of combinations between two symbolic constraints based on its logical operators.
Table~\ref{tbl:cost_function} describes the loss function for all forms of
symbolic constraints. Taking $a=b$ as an example, the loss function
$L=abs(a-b)$ achieves the minimum value $0$ when $a=b$, where $a$ and $b$ can
be arbitrary expressions. Thus, minimizing the loss function $L$ is equivalent
to solving the symbolic constraint. Similar logic is also useful to explain
the equivalence between the other kinds of symbolic constraints and its loss
function. These loss functions are not the only possible loss functions for these constraints. Any functions satisfying the Equation~\ref{eq:encode} can be used as loss functions and the same encoding mechanism can be applied to the other constraints.
Note that there are three special requirements for the encoding mechanism. 

\paragraph{Non-Zero Gradient Until SAT.}
The derivative of the loss function should not be zero until we find the
satisfiability results. For example, when we encode $a<b$, the derivative of
the loss function should not be equal to zero when $a=b$. Otherwise, the
NeuSolv will stop searching and return an unsatisfiable result. To guarantee that, we add
a small positive value $\alpha$ and adapts the loss function to be
$L=max(a-b+\alpha, 0)$ for the constraint $a<b$ and similarly for $a>b$ and $a
\neq b$. Taking the motivation sample shown in Section~\ref{sec:problem} as an example, the loss function is $L=max(99-ptr+0.5, 0)$ where $\alpha=0.5$

\paragraph{Fixed Lower Bound on Loss Function.}
The loss function for each constraint needs a fixed lower bound to avoid only minimizing the loss function of one constraint within the conjunctive constraints.
For instance, we should not encode $a \neq b$ to $L=-abs(a-b+\beta)$ as the loss
function can be negative infinity, where $\beta$ is a small real value. If the
constraint is $a \neq b \land c<d$ where $c$ and $d$ can be arbitrary
expressions, NeuSolv may only minimize the loss
function for $a \neq b$, because the loss function for $a \neq b \land c<d$ is the
sum of the loss function for $a \neq b$ and $c<d$. Thus, it may not find the
satisfiability result for both symbolic constraints. To avoid this, we add a lower bound
and adjust the loss function to be $L=max(-1, -abs(a-b+\beta))$. This lower bound ensures that the loss functions have a finite global minima.

\begin{table}[t]
\centering
\caption{Transforming symbolic constraints into the corresponding loss function. $a$ and $b$ represent arbitrary expressions. $S_1$ and $S_2$ represent arbitrary symbolic constraints. $L$ represents the loss function used for neural constraint solving. $L_{S_1}$ and $L_{S_2}$ represents the loss function for symbolic constraints $S_1$ and $S_2$ respectively. $\alpha$ represents a small positive value. $\beta$ represents a small real value.}
\label{tbl:cost_function}
\scalebox{0.9}{
\begin{tabular}{|l|l|}
\hline
\textbf{Symbolic Constraint} & \textbf{Loss Function ($L$)} \\ \hline
$S_1 \Coloneqq a<b$ & $L=max(a-b+\alpha, 0)$ \\ \hline
$S_1 \Coloneqq a>b$ & $L=max(b-a+\alpha, 0)$ \\ \hline
$S_1 \Coloneqq a \leq b$ & $L=max(a-b, 0)$ \\ \hline
$S_1 \Coloneqq a \geq b$ & $L=max(b-a, 0)$ \\ \hline
$S_1 \Coloneqq a=b$ & $L = abs(a-b)$ \\ \hline
$S_1 \Coloneqq a \neq b$ & $L = max(-1, -abs(a-b+\beta))$ \\ \hline
$S_1 \land S_2$ & $L = L_{S_1}+L_{S_2}$ \\ \hline
$S_1 \lor S_2$ & $L = min(L_{S_1}, L_{S_2})$ \\ \hline
\end{tabular}
}
\end{table}

\paragraph{Generality of Encoding.}
NeuSolv can only be applied to differentiable loss functions, because it requires computing the derivatives of the loss function.
Thus, \tool need to transform the expression $a$ and $b$ in Table~\ref{tbl:cost_function} to a differentiable function. The encoding mechanism of expressions is generic. As long as \tool can transform the expression into a differential function, any encoding mechanism can be plugged in \tool for neuro-symbolic constraint solving.

\subsection{Optimizations}~\label{sec:solve_impl}
\tool applies five optimization strategies to reduce the computation time for neuro-symbolic constraint solving.

\paragraph{Single Variable Update.} 
Given a set of input variables to neural constraint, \tool only updates one variable for each enumeration in NeuSolv. In order to select the variable, \tool computes the derivative values for each variable and sorts the absolute values of derivatives. The updated variable is the one with the largest absolute value of the derivative. 
This is because the derivative value for each element only computes the influence of changing the value of one variable towards the value of loss function, but does not measure the joint influence of multiple variables. Thus, updating them simultaneously may increase the loss value. Moreover, updating one variable per iteration allows the search engine to perform the minimum number of mutations on the initial input in order to prevent the input from being invalid.

\paragraph{Type-based Update.}
To ensure the input is valid, \tool adapts the update strategy according to the types of variables. If the variable is an integer, \tool first binarizes the value of derivatives and then updates the variables with the binarized value. If the variable is a float, \tool updates the variable with the actual derivatives.

\paragraph{Caching.}
\tool stores the updated results for each enumeration in NeuSolv. 
As the search algorithm is a deterministic approach, if we have the same input, neural constraints and the loss function, the final generated result is the same. 
Thus, to avoid unnecessary recomputation, \tool stores the update history and checks whether current input is cached in history. If yes, \tool reuses previous result; otherwise, \tool keeping searching for new input. 

\paragraph{SAT Checking Per Enumeration.}
To speed up the solving procedure, \tool verifies the satisfiability of the variables after each enumeration in NeuSolv. Once it satisfies the symbolic constraints, NeuSolv terminates and returns \texttt{SAT} to \tool. This is because not only the result achieving global minima can be the satisfiability result of symbolic constraint. For example, any result can be the satisfiability result of the constraint $a \neq b$ except for the result satisfying $a=b$. Hence, \tool does not wait for minimizing the loss function, but checks the updated result for every iteration. 

\paragraph{Parallelization.}
\tool executes NeuSolv with different initial input in parallel since each loop for solving mixed constraints is independent. This parallelization reduces the time for finding the global minima of the loss function. 

\section{Neural Constraint Learning}~\label{sec:learn_impl}
We have described the constraint solver for neuro-symbolic constraint solving; now it remains to discuss how \tool obtains the neural constraints. In this section, we discuss the design of neural constraint learning engine in \tool.

Given a program, the selection of network architecture is the key for learning any neural constraint. In this paper, we use multilayer perceptron (MLP) architecture which consists of multiple layers of nodes and connects each node with all nodes in the previous layer~\cite{rumelhart1985learning}. Each node in the same layer does not share any connections with others. We select this architecture because it is a suitable choice for the fixed-length inputs. There are other more efficient architectures (e.g., CNN~\cite{lawrence1997face, krizhevsky2012imagenet} and RNN~\cite{medsker2001recurrent,mikolov2010recurrent}) for the data with special relationships, and \tool gives users the flexibility to add more network architectures in \tool.

The selection of activation function plays significant role for neural constraint inference as well. In this paper, we consider multiple activation functions (e.g., Sigmoid and Tanh) and finally select the rectifier function Relu as the activation function, because Relu obtains parse representation and reduces the likelihood of vanishing gradient~\cite{glorot2011deep, maas2013rectifier}. In other words, the neural network with Relu has higher chance to converge than other activation functions. 

In addition, to ensure the generality of neural constraint, we implement an early-stopping mechanism which is a regularization approach to reduce over-fitting~\cite{yao2007early}. It stops the learning procedure when the current learned neural constraint behaves worse on unseen test executions than the previous constraint. As the unseen test executions are never used for learning the neural constraint, the performance of learned neural constraint on unseen test executions is a fair measure for the generality of learned neural constraints.

\tool can use any machine learning approach, optimization algorithm (e.g., momentum gradient descent~\cite{qian1999momentum} and AdaGrad~\cite{duchi2011adaptive}) and regularization solution (e.g., dropout~\cite{srivastava2014dropout} and Tikhonov regularization~\cite{tikhonov1963solution}) to learn the neural constraints.
With the advances in machine learning, \tool can adopt new architectures and learning approaches in neural constraint inference.



\section{Evaluation}

We implement \tool in Python and Google TensorFlow~\cite{tensorflow} with a total of 1808 lines of code for training the neural constraints and solving the neuro-symbolic constraints.
Our evaluation highlights two features of \tool: (a) it generates the exploits for 13/14 vulnerable programs; (b) it solves 100\% of the given neuro-symbolic constrains for each loop.

\paragraph{Experimental Setup.}
To evaluate \tool, we configure the maximum enumeration of NeuSolv $M_e$ to be 10000 after which NeuSolv will terminate. (discussed in Section~\ref{sec:design_overview}). The larger the maximum enumeration, the better the performance of \tool is on neural constraint solving.
Our experiments are performed on a server with 40-core Intel Xeon 2.6GHz CPUs with 64 GB of RAM. 

\subsection{Effectiveness in Exploit Generation}

\begin{table}[]
\centering
\scalebox{0.9}{
\begin{tabular}{|c|c|c|c|}
\hline
\textbf{Program} & \textbf{Vulnerable Condition} & \textbf{LD} & \textbf{\begin{tabular}[c]{@{}c@{}}Find\\ Exploits?\end{tabular}} \\ \hline
\textbf{BIND1} & $strlen(data)>4140$ & 16 & Yes \\ \hline
\textbf{BIND2} & $strlen(data)>4140$ & 12 & Yes \\ \hline
\textbf{BIND3} & $strlen(anbuf)>512$ & 13 & Yes \\ \hline
\textbf{BIND4} & $strlen(buf)>999$ & 52 & Yes \\ \hline
\textbf{Sendmail1} & $strlen(buf)>31$ & 1 & Yes \\ \hline
\textbf{Sendmail2} & $strlen(buf)>5$ & 38 & Yes \\ \hline
\textbf{Sendmail3} & $strlen(ooutfile)>50$ & 18 & Yes \\ \hline
\textbf{Sendmail4} & $strlen(fbuf)>51$ & 2 & Yes \\ \hline
\textbf{Sendmail5} & $strlen(pvpbuf)>50$ & 6 & Yes \\ \hline
\textbf{Sendmail6} & $strlen(tTvect)>100$ & 11 & No \\ \hline
\textbf{Sendmail7} & \begin{tabular}[c]{@{}c@{}}$strlen((*rr) \rightarrow$\\ $rr\_u.rr\_txt) > size$\end{tabular} & 16 & Yes \\ \hline
\textbf{WuFTP1} & $strlen(path)>10$ & 5 & Yes \\ \hline
\textbf{WuFTP2} & $strlen(resolved)>46$ & 29 & Yes \\ \hline
\textbf{WuFTP3} & $strlen(curpath)>46$ & 7 & Yes \\ \hline
\end{tabular}
}
\caption{\tool finds the exploits for 13 out of 14 programs in the buffer overflow benchmark. LD represents the number of branches of which the condition relies on loop counts but not input arguments, which indicates the complexity of the program for symbolic execution to analyze it.}
\label{tbl:lese_result}
\end{table}

To evaluate the effectiveness of \tool in exploit generation, we select 14 vulnerable programs with buffer overflows from open-source network servers (e.g., BIND, Sendmail and WuFTP)~\cite{LESE_benchmark}.
We choose this benchmark because it comprises of multiple loops and various complex control and data dependencies which are challenging for symbolic execution to handle (discussed in Section~\ref{sec:problem}). 
To measure the complexity of problems, we utilize the number of branches of which the condition is related to loop counts rather than input arguments in the vulnerable path. This metric is also used in ~\cite{LESE}. Table~\ref{tbl:lese_result} represents the complexity of each program along with the result of exploit generation.

To show the effectiveness of neuro-symbolic constraint learning and solving, for each program, we mark the code from the beginning of the program to the location accessing buffers to be represented as neural constraints. Then, we mark all inputs and all buffer lengths in the program as symbolic by default. 
In cases where we know the input format, we provide it as additional information in form of program annotations (for e.g., specific input field values).
In our example from Section~\ref{sec:problem}, to analyze the program which takes HTTP requests as input, \tool marks the uri and version field as well as the length of all the buffers as symbolic.
\tool randomly initializes the symbolic input arguments for each program, executes the  program and collects the values of variables of interest.  For our experiments,
 we collect up to $100000$ samples of such executions. $80$\% of these samples are used for learning the neural constraints, while remaining $20$\% are used for evaluating the accuracy of learned neural constraints. To get the vulnerable conditions, we manually analyze the source code and set it as symbolic constraint.

Using the above steps, our experiments show that \tool is able to find the correct exploit for 13 out of 14 programs in the
benchmark. Next, we compare the efficiency of \tool on buffer overflow exploit generation with an existing symbolic execution method called  Loop-Extended Symbolic Execution (LESE)~\cite{LESE} which is a dynamic symbolic execution based tool.
It is a heuristic-based approach which hard-codes the relationship between
loop counts and inputs. We reproduce LESE's machine configuration for fair comparison.
Our experiments show that \tool requires maximum two hours to find the exploits on this setup. On the other hand, LESE requires more than five hours.
Thus, \tool's performance is comparable to LESE for exploit generation.

In addition, the time that \tool spends in exploit generation is not dependent on the complexity of the target code, as \tool is a black-box approach for neural constraint learning. For example, the time spent for analyzing the program \texttt{Sendmail1} with one loop-dependent branch is as same as the time used for program \texttt{Sendmail3} with 18 loop-dependent branches.
 
\begin{framed}
\textbf{Finding 1:} \tool is able to find the correct exploit for 13 out of 14 programs.
\end{framed}

To check whether \tool learns the correct constraint, we manually analyze the weights of trained neural constraint (discussed in Appendix~\ref{sec:analyze_nn}). We find that \tool is able to learn the neural constraints representing the correct variable relationships. For example, in program \texttt{Sendmail7}, \tool not only learns that the final length of vulnerable buffer $(*$\texttt{rr}$)\rightarrow $\texttt{rr\_u.rr\_txt} is controlled by \texttt{txtlen} field of DNS response which is the $49^{th}$ element of the input, but also the fact that the allocated size for the vulnerable buffer is determined by the \texttt{size} field which is the $47^{th}$ and $48^{th}$ elements of DNS response. For the programs that \tool successfully generates exploits for, we manually analyze all the neural constraints and find that they all precisely represent the variable relationships in the source code. 

\begin{framed}
\textbf{Finding 2:} \tool learns the correct neural constraint to represent the variable relationships in the source code.
\end{framed}

\tool reaches timeout for exploit generation in only one program
(\texttt{Sendmail6}) where the buffer overflow is caused by the integer
overflow. \tool fails to generate exploits because the neural net treats integers as real values
and is not aware of the programmatic behavior that integers wrap around after
they exceed the maximum value representable by its bit width.  For example, to
capture 32-bit integer overflow, \tool needs to know the rule of integer
overflow where the value becomes negative if it is larger than
\texttt{0x7FFFFFFF} in x86. To address this, we can explicitly add this
rule as a part of symbolic constraints for all integer types and then
solve the neuro-symbolic constraints.

\subsection{Micro-Benchmarking of \tool}

\begin{figure}[]
\centering
\epsfig{file=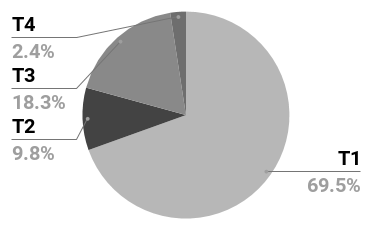, scale=0.4}
\caption{Type distribution of the symbolic constraints in NLA and HOLA. T1 represents the constraints with $\geq$ or $\leq$ operator; T2 presents the constraints with > or < operator; T3 represents the constraints with $==$ or $\neq$ operator; T4 represents the constraints with $\land$ or $\lor$ operators.}
\label{fig:type_distribution}
\end{figure}
\begin{table*}[]
\centering
\scalebox{0.9}{
\begin{tabular}{|c|c|c|c|c|c|c|c|c|c|c|c|c|c|c|c|}
\hline
P & Type & $T_{NS}$ & $T_{NE}$ & P & Type & $T_{NS}$ & $T_{NE}$ & P & Type & $T_{NS}$ & $T_{NE}$ & P & Type & $T_{NS}$ & $T_{NE}$ \\ \hline
cohendiv & T2 & 1 & 1 & dijkstra\_2 & T3 & 2 & - & prod4br & T4 & 1 & 1 & geo3 & T1 & 1 & 1 \\ \hline
divbin\_1 & T2 & 1 & 1 & freire1 & T1 & 1 & 1 & knuth & T4 & 1 & 1 & ps2 & T1 & 1 & 1 \\ \hline
divbin\_2 & T3 & 1 & 5 & freire2 & T1 & 1 & 1 & fermat1 & T3 & 1 & 1 & ps3 & T1 & 1 & 1 \\ \hline
mannadiv & T3 & 1 & 1 & cohencu & T2 & 1 & 1 & fermat2 & T3 & 3 & 3 & ps4 & T1 & 1 & 1 \\ \hline
hard\_1 & T2 & 1 & 1 & egcd & T3 & 1 & 2 & lcm1 & T3 & 1 & 1 & ps5 & T1 & 1 & 1 \\ \hline
hard\_2 & T3 & 1 & 5 & egcd2 & T3 & 1 & 1 & lcm2 & T3 & 1 & 4 & ps6 & T1 & 1 & 1 \\ \hline
sqrt1 & T1 & 1 & 1 & egcd3 & T3 & 1 & 5 & geo1 & T1 & 1 & 1 &  &  &  &  \\ \hline
dijkstra\_1 & T1 & 1 & 1 & prodbin & T3 & 1 & 1 & geo2 & T1 & 1 & 1 &  &  &  &  \\ \hline
\end{tabular}
}
\caption{Evaluation results of \tool's constraint solving on NLA benchmark. {\em P} represents the program name; {\em Type} shows the type of symbolic constraints; {\em $T_{NS}$} shows the number of trials NeuSolv takes for solving the given neuro-symbolic constraints; {\em $T_{NE}$} represents the number of trials \tool needs to reach the post-condition of the loop. {\em `-'} represents that \tool reaches its timeout before reaching the post-condition.}
\label{tbl:nla_result}
\end{table*}
\begin{table*}[]
\centering

\scalebox{0.9}{
\begin{tabular}{|c|c|c|c|c|c|c|c|c|c|c|c|c|c|c|c|}
\hline
P & Type & $T_{NS}$ & $T_{NE}$ & P & Type & $T_{NS}$ & $T_{NE}$ & P & Type & $T_{NS}$ & $T_{NE}$ & P & Type & $T_{NS}$ & $T_{NE}$ \\ \hline
01 & T1 & 1 & 1 & 12\_1 & T1 & 1 & 1 & 24 & T1 & 1 & 1 & 36\_2 & T1 & 1 & 1 \\ \hline
02 & T1 & 1 & 1 & 12\_2 & T2 & 1 & 1 & 25 & T1 & 1 & 1 & 37 & T1 & 1 & 1 \\ \hline
03 & T1 & 1 & 1 & 13 & T1 & 1 & 1 & 26 & T1 & 1 & 1 & 38 & T1 & 1 & 1 \\ \hline
04 & T1 & 1 & 2 & 14 & T2 & 3 & 3 & 27 & T1 & 1 & 1 & 39 & T3 & 1 & 1 \\ \hline
05 & T1 & 1 & 1 & 15 & T1 & 1 & 1 & 28\_1 & T1 & 1 & 1 & 40\_1 & T1 & 1 & 1 \\ \hline
06 & T1 & 1 & 1 & 16 & T3 & 1 & 2 & 28\_2 & T3 & 1 & 1 & 40\_2 & T1 & 1 & 1 \\ \hline
07 & T1 & 1 & 1 & 17 & T1 & 1 & 1 & 29 & T1 & 1 & 1 & 41 & T2 & 1 & 1 \\ \hline
08 & T1 & 1 & 1 & 18 & T1 & 1 & 1 & 31 & T1 & 1 & 1 & 42 & T1 & 1 & 1 \\ \hline
09\_1 & T1 & 1 & 1 & 19 & T1 & 1 & 1 & 32 & T1 & 1 & 1 & 43 & T1 & 1 & 1 \\ \hline
09\_2 & T1 & 1 & 1 & 20 & T1 & 1 & 1 & 33 & T1 & 1 & 1 & 44 & T2 & 2 & 2 \\ \hline
09\_3 & T1 & 1 & 1 & 21 & T1 & 1 & 1 & 34 & T1 & 1 & 1 & 45\_1 & T1 & 1 & 1 \\ \hline
09\_4 & T1 & 1 & 1 & 22 & T1 & 1 & 1 & 35 & T1 & 1 & 1 & 45\_2 & T1 & 1 & 1 \\ \hline
10 & T1 & 1 & 1 & 23 & T1 & 1 & 1 & 36\_1 & T1 & 1 & 1 & 46 & T1 & 1 & 1 \\ \hline
\end{tabular}
}
\caption{Evaluation results of \tool's constraint solving on HOLA benchmark. {\em P} represents the program name; {\em Type} shows the type of symbolic constraints; {\em $T_{NS}$} shows the number of trials NeuSolv takes for solving the given neuro-symbolic constraints; {\em $T_{NE}$} represents the number of trials \tool needs to reach the post-condition of the loop.}
\label{tbl:hola_result}
\end{table*}

We ask three empirical questions with our micro-benchmarks: 
\begin{enumerate}
\item How fast does \tool solve a given neuro-symbolic constraint? 
\item What is the accuracy of neural constraints learned by \tool? 
\item What is the influence of learning and solving on the overall efficiency of \tool?
\end{enumerate}


For this, we use two benchmarks, namely HOLA and NLA, which comprise $73$ programs
with $82$ loops and $259$ input variables in total. These two benchmarks are
widely used for invariant synthesis~\cite{DIG, DIG2,gupta2009invgen} which is
useful for formal verification. We select these two benchmarks because they
have various kinds of loop invariants and capturing them is known to be a
challenge for symbolic execution. To this end, we evaluate \tool's ability to
reach the post-condition of the loops in these benchmarks.
\begin{figure}
\begin{lstlisting}[style=JavaScript, xleftmargin=0.3cm, captionpos=b, language=C]
void func(int a, int b){
	int c,d,cnt; c = a; d = b;cnt=0;
	while(c>d){
		c = c+d+1; d = d+1;cnt++;
	}
}
\end{lstlisting}
\caption{A simple function with one loop.}
\label{fig:example}
\end{figure}
For each program, we mark the loop to be represented by neural constraints. In each loop, \tool needs to (1) learn the loop invariant $N$, (2) get the symbolic invariant of loop guard $S$ from the symbolic execution engine, and (3) solve $N \land \neg S$. Consider the example in Figure~\ref{fig:example}. 
\tool first learns the neural constraint $N: \{a,b,cnt\} \mapsto \{c,d\}$ representing the loop invariant on Line 5. Then, it gets the loop guard $c>d$ on Line 3 from the symbolic execution engine. Finally, it solves the
neuro-symbolic constraint $N \land c \leq d$. For each loop in our benchmarks, 
we mark all the input arguments (e.g., $a$ and $b$) as well as the loop count as symbolic.
If the loop count is not an explicit variable, \tool adds an implicit count incremented for each iteration to capture the number of iterations in the loop.
Figure~\ref{fig:type_distribution}
shows the type distribution of the negation of loop guards in NLA and HOLA
benchmarks which covers all kinds of constraints expressed in
Table~\ref{tbl:cost_function}.


\paragraph{Effectiveness of Neuro-Symbolic Constraint Solving.}
Recall that NeuSolv randomly sets an initial state when it begins  the
gradient-based optimization of a loss function.  If it fails to find a
satisfiability result before the timeout, \tool needs to restart the
search from a different initial state  because 
the search is dependent on the initial state
(discussed in Section~\ref{sec:design_overview}). We call each search attempt from a
new initial state as one {\em trial}. Thus, to evaluate how fast \tool solves
a given neuro-symbolic constraint, we  use the number of trials
that NeuSolv takes as the metric. The lower the number of trials that \tool
needs, the faster the neuro-symbolic constraint solving.  $T_{NS}$ column in
Table~\ref{tbl:nla_result} and Table~\ref{tbl:hola_result} shows the number of
trials \tool required to solve the given neuro-symbolic constraints for
each loop in NLA and HOLA benchmarks. From these results, we find that \tool
successfully solves $100\%$ of the given neuro-symbolic constraints with a
maximum of three trials. Among $82$ loops, \tool solves $95\%$ of 
neuro-symbolic constraints with only one trial. This result indicates that \tool can
successfully solve various kinds of given neuro-symbolic constraints
efficiently.

\begin{framed}
\textbf{Finding 3:} \tool is effective in neuro-symbolic constraint solving for 100\% of constraints with a maximum of three trials.
\end{framed}

\tool needs more than one trials for $4 / 82$ loops because of two main
reasons. First, our current timeout value is not enough for solving the
constraints in two cases (program $14$ and $40$ in HOLA benchmark). To address
this, we can either increase the timeout or restart  the search with a new
initial state.  We experiment on both options and report that the
latter can solve the constraints faster. For example, in program $40$, \tool
solves the given neuro-symbolic constraints within $2$ trials, but it reaches timeout
for one trial where the timeout is increased to three-folds. For the remaining
two loops, \tool fails because of the inefficiency  of gradient-based search
in NeuSolv. For example, in program \texttt{fermat2}, NeuSolv gets stuck at
the saddle point. To address this, we can apply trust region
algorithm~\cite{sorensen1982newton} or cubic
regularization~\cite{nesterov2006cubic} which utilizes second-order derivative
to find and avoid saddle points.

\paragraph{Accuracy of Neural Constraint Learning.}
To measure the effectiveness of neural constraint learning, we computes the learning accuracy $Acc$ which is defined as:
$
Acc = \frac{M_R}{M}
$,
where $M_R$ is the number of (unseen) test executions where learned neural constraints predict the right outputs and $M$ is the total tested executions. The higher the accuracy, the more precise the learned neural constraints. For $82$ loops in our benchmarks,
\tool achieves more than $80\%$ accuracy for $66$ neural constraints. For
example, \tool achieves $97\%$ accuracy for learning the second loop invariant
in program $hard$ which contains multiple multiplications with divisions.

\begin{framed}
\textbf{Finding 4:} \tool achieves more than $80\%$ learning accuracy for $66/82$ neural constraints.
\end{framed}

\paragraph{Combined (Learning + Solving) Efficiency.}
There are two steps involved in solving a neuro-symbolic task (reaching the
post-condition in this case) namely: infer the constraints and solve them.  So
far in our micro-benchmarks, we have evaluated these two steps independent of
each other.
For completeness, we now present our experimental analysis for understanding how these two steps affect the  overall efficiency of \tool in performing a
given task. 

\begin{table}[]
\centering
\scalebox{1}{
\begin{tabular}{|c|c|c|}
\hline
\backslashbox{\textbf{Solving}}{\textbf{Learning}} & \textbf{Success} & \textbf{Failure} \\ \hline
\textbf{Success}                                             & 61/63            & 10/15            \\ \hline
\textbf{Failure}                                             & 0/3              & 0/1              \\ \hline
\end{tabular}
}
\caption{
Effect of constraint learning and solving  on \tool's overall efficiency.
We classify constraint learning to be a success when accuracy $\geq 80\%$ and a failure otherwise.
We classify constraint solving to be a success when \tool solves the given constraints with one trial and a failure otherwise.
We classify task solving to be a success when the concrete values generated with 1 trial reaches the post-condition and a failure otherwise.
The cell value represents the number of loops which succeed in task solving out of total loops under that category.
}
\label{tbl:end}
\end{table}

\tool successfully solves $71$ out of $82$ end-to-end tasks in total.  
Table~\ref{tbl:end} shows the contributions of each step in solving a neuro-symbolic
task. When both steps are successful, \tool succeeds in solving $96.8\%$ of tasks (in top left cell),
However, when only \tool's solving is unsuccessful ($4$ cases in bottom left cell),  it always fails to complete the task. This shows that task solving is directly
dependent on  constraint solving, and justifies our focus on improving the
neuro-symbolic constraint solving efficiency in our constraint solver.
Ideally, \tool must always learn the constraints accurately as well as  solve
the constraints successfully in order to guarantee post-condition
reachability.  However, we notice that even when learning is inaccurate,
\tool is still able to solve the $66.7\%$ of the tasks (in top right cell). This
is because  \tool is at least able to learn the {\em trend} of certain variables
involved in the constraints if not the precise constraints.  Consider the
example in Figure~\ref{fig:example}. If the neural constraint learns
$c=a+4 \times cnt^2 \land d=b+cnt$, \tool finds the satisfiability result $a=2$,
$b=5$, $cnt=1$, $c=6$ and $d=6$. Even though the neural constraint does not
capture the precise loop invariant $c=a+cnt \times \frac{2 \times b+1+cnt}{2} \land
d=b+cnt$, it at least knows that the value of $c$ increases with the increase
in $cnt^2$. This partial learning aids \tool to solve the task and find $a=2$,
$b=5$ and $cnt=1$. Thus, we conclude that although learning is important, it
does not affect task solving as drastically as constraint solving. This
highlights the importance of effectiveness in constraint solving.

\begin{framed}
\textbf{Finding 5:} Constraint solving affects \tool's effectiveness more significantly than constraint learning.
\end{framed}

\section{Related Work}
\label{sec:related}
\tool is a new design point in constraint synthesis and constraint solving. In this section, we discuss the problems of the existing symbolic execution tools to show how \tool can handle it and presents how \tool differs from existing constraint synthesis.

\subsection{Symbolic Execution}
Symbolic execution \cite{King} has been used for program verification \cite{Dannenberg}, software testing \cite{King,klee}, and program repair via specification inference \cite{nguyen2013semfix}. 
In the last decade, we have witnessed an increased adoption of dynamic symbolic execution \cite{DART} where symbolic execution is used to partition of the input space, with the goal of achieving increased behavioral coverage. The input partitions computed are often defined as program paths, all inputs tracing the same path belong to the same partition. Thus, the test generation achieved by dynamic symbolic execution suffers from the path explosion problem. The problem of path explosion can be exacerbated owing to the presence of complex control flows, including long-running loops (which may affect the scalability of dynamic symbolic execution since it involves loop unrolling) and external libraries. However, \tool does not suffer from the path explosion as it learns the constraints from test executions directly. 

Tackling path explosion is a major challenge in symbolic execution. Boonstopel et al. suggest the pruning of redundant paths during the symbolic execution tree construction~\cite{Boonstoppel}. Veritesting alternates between dynamic symbolic execution and static symbolic execution to mitigate path explosion~\cite{avgerinos2014enhancing}. The other predominant way of tackling the path explosion problem is by summarizing the behavior of code fragments in a program~\cite{godefroid,Anand,DSM-kuznetsov,veritesting,PESO,multiSE}. Simply speaking, a summarization technique provides an approximation of the behavior of certain fragments of a program to keep the scalability of symbolic execution manageable. Such an approximation of behaviors is also useful when certain code fragments, such as remote calls and libraries written in a different language, are not available for analysis. 

Among the past approaches supporting approximation of behaviors of (parts of) a program, the use of function summaries have been studied by Godefroid \cite{godefroid}. Such function summaries can also be computed on-demand \cite{Anand}.
Kuznetsov et al. present a selective technique to merge dynamic states. It merges two dynamic symbolic execution runs based on an estimation of the difficulty in solving the resultant Satisfiability Modulo Theory (SMT) constraints~\cite{DSM-kuznetsov}. Veritesting suggests supporting dynamic symbolic execution with static symbolic execution thereby alleviating path explosion due to factors such as loop unrolling\cite{veritesting}. 
The works of \cite{PESO,multiSE} suggest grouping together paths based on similar symbolic expressions in variables, and use such symbolic expressions as dynamic summaries to group paths.

\subsection{Constraints Synthesis}
To support the summarization of program behaviors, the other core technical primitive we can use is constraint synthesis. In our work, we propose a new constraint synthesis approach which utilizes neural networks to learn the constraints which are infeasible for symbolic execution. In comparison with previous solutions, the major difference is that \tool does not require any pre-defined templates of constraints and can learn any kind of relationships between variables. 

Over the last decade, there are two lines of works in constraint synthesis: white-box and black-box approaches. White-box constraint inference relies on a combination of light-weight techniques such as abstract interpretation~\cite{beyer2007path, cousot1977abstract, cousot1978automatic, cousot2005astree, mine2004weakly, rodriguez2007generating, rodriguez2007automatic}, interpolation~\cite{chen2015counterexample,mcmillan2013interpolation,jhala2006practical} or model checking algorithm IC3~\cite{bradley2011sat}. 
Although some white-box approaches can provide sound and complete constraints~\cite{colon2003linear}, it is dependent on the availability of source code and a human-specified semantics of the source language. Constructing these tools have required considerable manual expertise to achieve precision, and many of these techniques can be highly computationally intensive. 

To handle the unavailability of source code, there also exist a rich class of works on reverse engineering from dynamic executions~\cite{gupta2009invgen,garg2014ice,Daikon, sankaranarayanan2008dynamic, DIG, DIG2,padhi2017data}. 
Such works can be used to generate summaries of observed behavior from test executions. These summaries are not guaranteed to be complete. On the other hand, such incomplete summaries can be obtained from tests, and hence the source code of the code fragment being summarized need not be available. Daikon \cite{Daikon} is one of the earlier works proposing synthesis of potential invariants from values observed in test executions. The invariants supported in Daikon are in the form of linear relations among program variables. DIG extends Daikon to enable dynamic discovery of non-linear polynomial invariants via a combination of techniques including equation solving and polyhedral reasoning~\cite{DIG}. Krishna et al. use the decision tree, a machine learning technique, to learn the inductive constraints from good and bad test executions~\cite{krishna2015learning}. 

\tool devises a new gradient-based constraint solver which is the first work to support the solving of the conjunction of neural and SMT constraints. A similar gradient-based approach is also used in Angora~\cite{chen2018angora}, albeit for a completely different usage. It treats the predicates of branches as a black-box function which is not differentiable. Then, it computes the changes on the predicates by directly mutating the values of each variable in order to find the direction for changing variables. Similarly, Li et al. utilize the number of satisfied primitive constraints in a path condition as the target function for optimization and applies RACOS algorithm~\cite{yu2016derivative} to optimize the non-differentiable function for complementing symbolic execution~\cite{li2016symbolic}. However, \tool learns a differentiable function to represent the behaviors of the program from the test cases, encodes the symbolic constraints into a differentiable function and embeds it into neural constraints. It computes the values of derivatives for each variable for updating.


A recent work \cite{NS18} suggests the combination of neural reasoning and symbolic reasoning, albeit for an entirely different purpose, automated repair of student programming assignments. In contrast, our proposed neuro-symbolic execution solves neural and symbolic constraints together, and can be seen as a general purpose testing and analysis engine for programs.

\section{Conclusions}
To our knowledge, \tool is the first work utilizing neural networks to learn the constraints from values observed in test executions without pre-defined templates. \tool offers a new design point to simultaneously solve both symbolic constraints and neural constraints effectively, which can be used for complementing symbolic execution. It achieves good performance in both neuro-symbolic constraint solving and exploit generation for buffer overflows.
\begin{acks}
We thank Marcel B{\"o}hme for participating in the initial discussion of the project. We also thank Shruti Tople, Shin Hwei Tan and Xiang Gao for useful feedback on earlier drafts of this paper.  
This research is supported by a research grant from DSO, Singapore. All opinions expressed in this paper are solely those of the authors.
\end{acks}

\bibliographystyle{ACM-Reference-Format}
\balance
\bibliography{paper}


\begin{thebibliography}{00}


\ifx \showCODEN    \undefined \def \showCODEN     #1{\unskip}     \fi
\ifx \showDOI      \undefined \def \showDOI       #1{#1}\fi
\ifx \showISBNx    \undefined \def \showISBNx     #1{\unskip}     \fi
\ifx \showISBNxiii \undefined \def \showISBNxiii  #1{\unskip}     \fi
\ifx \showISSN     \undefined \def \showISSN      #1{\unskip}     \fi
\ifx \showLCCN     \undefined \def \showLCCN      #1{\unskip}     \fi
\ifx \shownote     \undefined \def \shownote      #1{#1}          \fi
\ifx \showarticletitle \undefined \def \showarticletitle #1{#1}   \fi
\ifx \showURL      \undefined \def \showURL       {\relax}        \fi
\providecommand\bibfield[2]{#2}
\providecommand\bibinfo[2]{#2}
\providecommand\natexlab[1]{#1}
\providecommand\showeprint[2][]{arXiv:#2}

\bibitem[\protect\citeauthoryear{??}{jal}{2018}]%
        {jalangi2}
 \bibinfo{year}{2018}\natexlab{}.
\newblock \bibinfo{title}{{Jalangi2: Dynamic analysis framework for
  JavaScript}}.
\newblock \bibinfo{howpublished}{\url{https://github.com/Samsung/jalangi2}}.
  (\bibinfo{year}{2018}).
\newblock


\bibitem[\protect\citeauthoryear{??}{PyE}{2018}]%
        {PyExZ3}
 \bibinfo{year}{2018}\natexlab{}.
\newblock \bibinfo{title}{{PyExZ3: Python Exploration with Z3}}.
\newblock \bibinfo{howpublished}{\url{https://github.com/thomasjball/PyExZ3}}.
   (\bibinfo{year}{2018}).
\newblock


\bibitem[\protect\citeauthoryear{Abadi, Barham, Chen, Chen, Davis, Dean, Devin,
  Ghemawat, Irving, Isard, et~al\mbox{.}}{Abadi et~al\mbox{.}}{2016}]%
        {tensorflow}
\bibfield{author}{\bibinfo{person}{Mart{\'\i}n Abadi}, \bibinfo{person}{Paul
  Barham}, \bibinfo{person}{Jianmin Chen}, \bibinfo{person}{Zhifeng Chen},
  \bibinfo{person}{Andy Davis}, \bibinfo{person}{Jeffrey Dean},
  \bibinfo{person}{Matthieu Devin}, \bibinfo{person}{Sanjay Ghemawat},
  \bibinfo{person}{Geoffrey Irving}, \bibinfo{person}{Michael Isard},
  {et~al\mbox{.}}} \bibinfo{year}{2016}\natexlab{}.
\newblock \showarticletitle{TensorFlow: A System for Large-Scale Machine
  Learning.}. In \bibinfo{booktitle}{{\em OSDI}}, Vol.~\bibinfo{volume}{16}.
  \bibinfo{pages}{265--283}.
\newblock


\bibitem[\protect\citeauthoryear{{\'A}brah{\'a}m}{{\'A}brah{\'a}m}{2015}]%
        {abraham2015building}
\bibfield{author}{\bibinfo{person}{Erika {\'A}brah{\'a}m}.}
  \bibinfo{year}{2015}\natexlab{}.
\newblock \showarticletitle{Building bridges between symbolic computation and
  satisfiability checking}. In \bibinfo{booktitle}{{\em Proceedings of the 2015
  ACM on International Symposium on Symbolic and Algebraic Computation}}. ACM,
  \bibinfo{pages}{1--6}.
\newblock


\bibitem[\protect\citeauthoryear{Anand, Godefroid, and Tillman}{Anand
  et~al\mbox{.}}{2008}]%
        {Anand}
\bibfield{author}{\bibinfo{person}{S. Anand}, \bibinfo{person}{P. Godefroid},
  {and} \bibinfo{person}{N. Tillman}.} \bibinfo{year}{2008}\natexlab{}.
\newblock \showarticletitle{Demand driven compositional symbolic execution}. In
  \bibinfo{booktitle}{{\em International Conference on Tools and Algortihms for
  Construction and Analysis of Systems ({TACAS})}}.
\newblock


\bibitem[\protect\citeauthoryear{Anand, Orso, and Harrold}{Anand
  et~al\mbox{.}}{2007}]%
        {anand2007type}
\bibfield{author}{\bibinfo{person}{Saswat Anand}, \bibinfo{person}{Alessandro
  Orso}, {and} \bibinfo{person}{Mary~Jean Harrold}.}
  \bibinfo{year}{2007}\natexlab{}.
\newblock \showarticletitle{Type-dependence analysis and program transformation
  for symbolic execution}. In \bibinfo{booktitle}{{\em International Conference
  on Tools and Algorithms for the Construction and Analysis of Systems}}.
  Springer, \bibinfo{pages}{117--133}.
\newblock


\bibitem[\protect\citeauthoryear{Andoni, Panigrahy, Valiant, and Zhang}{Andoni
  et~al\mbox{.}}{2014}]%
        {andoni2014learning}
\bibfield{author}{\bibinfo{person}{Alexandr Andoni}, \bibinfo{person}{Rina
  Panigrahy}, \bibinfo{person}{Gregory Valiant}, {and} \bibinfo{person}{Li
  Zhang}.} \bibinfo{year}{2014}\natexlab{}.
\newblock \showarticletitle{Learning polynomials with neural networks}. In
  \bibinfo{booktitle}{{\em International Conference on Machine Learning}}.
  \bibinfo{pages}{1908--1916}.
\newblock


\bibitem[\protect\citeauthoryear{Avgerinos, Rebert, Cha, and Brumley}{Avgerinos
  et~al\mbox{.}}{2014a}]%
        {veritesting}
\bibfield{author}{\bibinfo{person}{T. Avgerinos}, \bibinfo{person}{A. Rebert},
  \bibinfo{person}{S.K. Cha}, {and} \bibinfo{person}{D. Brumley}.}
  \bibinfo{year}{2014}\natexlab{a}.
\newblock \showarticletitle{Enhancing Symbolic Execution with Veritesting}. In
  \bibinfo{booktitle}{{\em Proceedings of International Conference on Software
  Engineering ({ICSE})}}.
\newblock


\bibitem[\protect\citeauthoryear{Avgerinos, Rebert, Cha, and Brumley}{Avgerinos
  et~al\mbox{.}}{2014b}]%
        {avgerinos2014enhancing}
\bibfield{author}{\bibinfo{person}{Thanassis Avgerinos},
  \bibinfo{person}{Alexandre Rebert}, \bibinfo{person}{Sang~Kil Cha}, {and}
  \bibinfo{person}{David Brumley}.} \bibinfo{year}{2014}\natexlab{b}.
\newblock \showarticletitle{Enhancing symbolic execution with veritesting}. In
  \bibinfo{booktitle}{{\em Proceedings of the 36th International Conference on
  Software Engineering}}. ACM, \bibinfo{pages}{1083--1094}.
\newblock


\bibitem[\protect\citeauthoryear{Baldoni, Coppa, D'Elia, Demetrescu, and
  Finocchi}{Baldoni et~al\mbox{.}}{2018}]%
        {SurveySymExec}
\bibfield{author}{\bibinfo{person}{Roberto Baldoni}, \bibinfo{person}{Emilio
  Coppa}, \bibinfo{person}{Daniele~Cono D'Elia}, \bibinfo{person}{Camil
  Demetrescu}, {and} \bibinfo{person}{Irene Finocchi}.}
  \bibinfo{year}{2018}\natexlab{}.
\newblock \showarticletitle{A Survey of Symbolic Execution Techniques}.
\newblock \bibinfo{journal}{{\em ACM Comput. Surv.\/}} \bibinfo{volume}{51},
  \bibinfo{number}{3}, Article \bibinfo{articleno}{50} (\bibinfo{year}{2018}).
\newblock


\bibitem[\protect\citeauthoryear{Beyer, Henzinger, Majumdar, and
  Rybalchenko}{Beyer et~al\mbox{.}}{2007}]%
        {beyer2007path}
\bibfield{author}{\bibinfo{person}{Dirk Beyer}, \bibinfo{person}{Thomas~A
  Henzinger}, \bibinfo{person}{Rupak Majumdar}, {and} \bibinfo{person}{Andrey
  Rybalchenko}.} \bibinfo{year}{2007}\natexlab{}.
\newblock \showarticletitle{Path invariants}. In \bibinfo{booktitle}{{\em Acm
  Sigplan Notices}}, Vol.~\bibinfo{volume}{42}. ACM, \bibinfo{pages}{300--309}.
\newblock


\bibitem[\protect\citeauthoryear{Bhatia, Kohli, and Singh}{Bhatia
  et~al\mbox{.}}{2018}]%
        {NS18}
\bibfield{author}{\bibinfo{person}{S. Bhatia}, \bibinfo{person}{P. Kohli},
  {and} \bibinfo{person}{R. Singh}.} \bibinfo{year}{2018}\natexlab{}.
\newblock \showarticletitle{Neuro-Symbolic Program Corrector for Introductory
  Programming Assignments}. In \bibinfo{booktitle}{{\em International
  Conference on Software Engineering ({ICSE})}}.
\newblock


\bibitem[\protect\citeauthoryear{Bojarski, Del~Testa, Dworakowski, Firner,
  Flepp, Goyal, Jackel, Monfort, Muller, Zhang, et~al\mbox{.}}{Bojarski
  et~al\mbox{.}}{2016}]%
        {bojarski2016end}
\bibfield{author}{\bibinfo{person}{Mariusz Bojarski}, \bibinfo{person}{Davide
  Del~Testa}, \bibinfo{person}{Daniel Dworakowski}, \bibinfo{person}{Bernhard
  Firner}, \bibinfo{person}{Beat Flepp}, \bibinfo{person}{Prasoon Goyal},
  \bibinfo{person}{Lawrence~D Jackel}, \bibinfo{person}{Mathew Monfort},
  \bibinfo{person}{Urs Muller}, \bibinfo{person}{Jiakai Zhang},
  {et~al\mbox{.}}} \bibinfo{year}{2016}\natexlab{}.
\newblock \showarticletitle{End to end learning for self-driving cars}.
\newblock \bibinfo{journal}{{\em arXiv preprint arXiv:1604.07316\/}}
  (\bibinfo{year}{2016}).
\newblock


\bibitem[\protect\citeauthoryear{Boonstoppel, Cadar, and Engler}{Boonstoppel
  et~al\mbox{.}}{2008}]%
        {Boonstoppel}
\bibfield{author}{\bibinfo{person}{P. Boonstoppel}, \bibinfo{person}{C. Cadar},
  {and} \bibinfo{person}{D. Engler}.} \bibinfo{year}{2008}\natexlab{}.
\newblock \showarticletitle{RWset: Attacking path explosion in constraint-based
  test generation}. In \bibinfo{booktitle}{{\em International Conference on
  Tools and Algortihms for Construction and Analysis of Systems ({TACAS})}}.
\newblock


\bibitem[\protect\citeauthoryear{Bradley}{Bradley}{2011}]%
        {bradley2011sat}
\bibfield{author}{\bibinfo{person}{Aaron~R Bradley}.}
  \bibinfo{year}{2011}\natexlab{}.
\newblock \showarticletitle{SAT-based model checking without unrolling}. In
  \bibinfo{booktitle}{{\em International Workshop on Verification, Model
  Checking, and Abstract Interpretation}}. Springer, \bibinfo{pages}{70--87}.
\newblock


\bibitem[\protect\citeauthoryear{Bundy and Wallen}{Bundy and Wallen}{1984}]%
        {bundy1984breadth}
\bibfield{author}{\bibinfo{person}{Alan Bundy} {and} \bibinfo{person}{Lincoln
  Wallen}.} \bibinfo{year}{1984}\natexlab{}.
\newblock \showarticletitle{Breadth-first search}.
\newblock In \bibinfo{booktitle}{{\em Catalogue of Artificial Intelligence
  Tools}}. \bibinfo{publisher}{Springer}, \bibinfo{pages}{13--13}.
\newblock


\bibitem[\protect\citeauthoryear{Cadar, Dunbar, Engler, et~al\mbox{.}}{Cadar
  et~al\mbox{.}}{2008}]%
        {klee}
\bibfield{author}{\bibinfo{person}{Cristian Cadar}, \bibinfo{person}{Daniel
  Dunbar}, \bibinfo{person}{Dawson~R Engler}, {et~al\mbox{.}}}
  \bibinfo{year}{2008}\natexlab{}.
\newblock \showarticletitle{KLEE: Unassisted and Automatic Generation of
  High-Coverage Tests for Complex Systems Programs.}
\newblock \bibinfo{journal}{{\em Proceedings of the USENIX Symposium on
  Operating System Design and Implementation\/}}  \bibinfo{volume}{8},
  \bibinfo{pages}{209--224}.
\newblock


\bibitem[\protect\citeauthoryear{Cadar, Ganesh, Pawlowski, Dill, and
  Engler}{Cadar et~al\mbox{.}}{2006}]%
        {exe}
\bibfield{author}{\bibinfo{person}{Cristian Cadar}, \bibinfo{person}{Vijay
  Ganesh}, \bibinfo{person}{Peter~M. Pawlowski}, \bibinfo{person}{David~L.
  Dill}, {and} \bibinfo{person}{Dawson~R. Engler}.}
  \bibinfo{year}{2006}\natexlab{}.
\newblock \showarticletitle{EXE: Automatically Generating Inputs of Death}. In
  \bibinfo{booktitle}{{\em Proceedings of the 13th ACM Conference on Computer
  and Communications Security}} {\em (\bibinfo{series}{CCS '06})}.
  \bibinfo{publisher}{ACM}, \bibinfo{address}{New York, NY, USA},
  \bibinfo{pages}{322--335}.
\newblock
\showISBNx{1-59593-518-5}
\showDOI{%
\url{https://doi.org/10.1145/1180405.1180445}}


\bibitem[\protect\citeauthoryear{Cadar and Sen}{Cadar and Sen}{2013}]%
        {cadar2013symbolic}
\bibfield{author}{\bibinfo{person}{Cristian Cadar} {and}
  \bibinfo{person}{Koushik Sen}.} \bibinfo{year}{2013}\natexlab{}.
\newblock \showarticletitle{Symbolic execution for software testing: three
  decades later}.
\newblock \bibinfo{journal}{{\it Commun. ACM}} \bibinfo{volume}{56},
  \bibinfo{number}{2} (\bibinfo{year}{2013}), \bibinfo{pages}{82--90}.
\newblock


\bibitem[\protect\citeauthoryear{Canini, Venzano, Peresini, Kostic, and
  Rexford}{Canini et~al\mbox{.}}{2012}]%
        {canini2012nice}
\bibfield{author}{\bibinfo{person}{Marco Canini}, \bibinfo{person}{Daniele
  Venzano}, \bibinfo{person}{Peter Peresini}, \bibinfo{person}{Dejan Kostic},
  {and} \bibinfo{person}{Jennifer Rexford}.} \bibinfo{year}{2012}\natexlab{}.
\newblock \showarticletitle{A NICE way to test OpenFlow applications}. In
  \bibinfo{booktitle}{{\em Proceedings of the 9th USENIX Symposium on Networked
  Systems Design and Implementation (NSDI)}}.
\newblock


\bibitem[\protect\citeauthoryear{Chen and Chen}{Chen and Chen}{2018}]%
        {chen2018angora}
\bibfield{author}{\bibinfo{person}{Peng Chen} {and} \bibinfo{person}{Hao
  Chen}.} \bibinfo{year}{2018}\natexlab{}.
\newblock \showarticletitle{Angora: Efficient Fuzzing by Principled Search}.
\newblock \bibinfo{journal}{{\em arXiv preprint arXiv:1803.01307\/}}
  (\bibinfo{year}{2018}).
\newblock


\bibitem[\protect\citeauthoryear{Chen, Hong, Wang, and Zhang}{Chen
  et~al\mbox{.}}{2015}]%
        {chen2015counterexample}
\bibfield{author}{\bibinfo{person}{Yu-Fang Chen}, \bibinfo{person}{Chih-Duo
  Hong}, \bibinfo{person}{Bow-Yaw Wang}, {and} \bibinfo{person}{Lijun Zhang}.}
  \bibinfo{year}{2015}\natexlab{}.
\newblock \showarticletitle{Counterexample-guided polynomial loop invariant
  generation by lagrange interpolation}. In \bibinfo{booktitle}{{\em
  International Conference on Computer Aided Verification}}. Springer,
  \bibinfo{pages}{658--674}.
\newblock


\bibitem[\protect\citeauthoryear{Chipounov, Kuznetsov, and Candea}{Chipounov
  et~al\mbox{.}}{2011}]%
        {chipounov2011s2e}
\bibfield{author}{\bibinfo{person}{Vitaly Chipounov},
  \bibinfo{person}{Volodymyr Kuznetsov}, {and} \bibinfo{person}{George
  Candea}.} \bibinfo{year}{2011}\natexlab{}.
\newblock \showarticletitle{S2E: A platform for in-vivo multi-path analysis of
  software systems}.
\newblock \bibinfo{journal}{{\em ACM SIGPLAN Notices\/}} \bibinfo{volume}{46},
  \bibinfo{number}{3} (\bibinfo{year}{2011}), \bibinfo{pages}{265--278}.
\newblock


\bibitem[\protect\citeauthoryear{Coen-Porisini, Denaro, Ghezzi, and
  Pezz{\'e}}{Coen-Porisini et~al\mbox{.}}{2001}]%
        {coen2001using}
\bibfield{author}{\bibinfo{person}{Alberto Coen-Porisini},
  \bibinfo{person}{Giovanni Denaro}, \bibinfo{person}{Carlo Ghezzi}, {and}
  \bibinfo{person}{Mauro Pezz{\'e}}.} \bibinfo{year}{2001}\natexlab{}.
\newblock \showarticletitle{Using symbolic execution for verifying
  safety-critical systems}. In \bibinfo{booktitle}{{\em ACM SIGSOFT Software
  Engineering Notes}}, Vol.~\bibinfo{volume}{26}. ACM,
  \bibinfo{pages}{142--151}.
\newblock


\bibitem[\protect\citeauthoryear{Col{\'o}n, Sankaranarayanan, and
  Sipma}{Col{\'o}n et~al\mbox{.}}{2003}]%
        {colon2003linear}
\bibfield{author}{\bibinfo{person}{Michael~A Col{\'o}n},
  \bibinfo{person}{Sriram Sankaranarayanan}, {and} \bibinfo{person}{Henny~B
  Sipma}.} \bibinfo{year}{2003}\natexlab{}.
\newblock \showarticletitle{Linear invariant generation using non-linear
  constraint solving}. In \bibinfo{booktitle}{{\em International Conference on
  Computer Aided Verification}}. Springer, \bibinfo{pages}{420--432}.
\newblock


\bibitem[\protect\citeauthoryear{Cousot and Cousot}{Cousot and Cousot}{1977}]%
        {cousot1977abstract}
\bibfield{author}{\bibinfo{person}{Patrick Cousot} {and}
  \bibinfo{person}{Radhia Cousot}.} \bibinfo{year}{1977}\natexlab{}.
\newblock \showarticletitle{Abstract interpretation: a unified lattice model
  for static analysis of programs by construction or approximation of
  fixpoints}. In \bibinfo{booktitle}{{\em Proceedings of the 4th ACM
  SIGACT-SIGPLAN symposium on Principles of programming languages}}. ACM,
  \bibinfo{pages}{238--252}.
\newblock


\bibitem[\protect\citeauthoryear{Cousot, Cousot, Feret, Mauborgne, Min{\'e},
  Monniaux, and Rival}{Cousot et~al\mbox{.}}{2005}]%
        {cousot2005astree}
\bibfield{author}{\bibinfo{person}{Patrick Cousot}, \bibinfo{person}{Radhia
  Cousot}, \bibinfo{person}{J{\'e}r{\^o}me Feret}, \bibinfo{person}{Laurent
  Mauborgne}, \bibinfo{person}{Antoine Min{\'e}}, \bibinfo{person}{David
  Monniaux}, {and} \bibinfo{person}{Xavier Rival}.}
  \bibinfo{year}{2005}\natexlab{}.
\newblock \showarticletitle{The ASTR{\'E}E analyzer}. In
  \bibinfo{booktitle}{{\em European Symposium on Programming}}. Springer,
  \bibinfo{pages}{21--30}.
\newblock


\bibitem[\protect\citeauthoryear{Cousot and Halbwachs}{Cousot and
  Halbwachs}{1978}]%
        {cousot1978automatic}
\bibfield{author}{\bibinfo{person}{Patrick Cousot} {and}
  \bibinfo{person}{Nicolas Halbwachs}.} \bibinfo{year}{1978}\natexlab{}.
\newblock \showarticletitle{Automatic discovery of linear restraints among
  variables of a program}. In \bibinfo{booktitle}{{\em Proceedings of the 5th
  ACM SIGACT-SIGPLAN symposium on Principles of programming languages}}. ACM,
  \bibinfo{pages}{84--96}.
\newblock


\bibitem[\protect\citeauthoryear{Cui, Kannan, and Wang}{Cui
  et~al\mbox{.}}{2007}]%
        {cui2007discoverer}
\bibfield{author}{\bibinfo{person}{Weidong Cui}, \bibinfo{person}{Jayanthkumar
  Kannan}, {and} \bibinfo{person}{Helen~J Wang}.}
  \bibinfo{year}{2007}\natexlab{}.
\newblock \showarticletitle{Discoverer: Automatic Protocol Reverse Engineering
  from Network Traces.}. In \bibinfo{booktitle}{{\em USENIX Security
  Symposium}}. \bibinfo{pages}{1--14}.
\newblock


\bibitem[\protect\citeauthoryear{Daniel, Gvero, and Marinov}{Daniel
  et~al\mbox{.}}{2010}]%
        {daniel2010test}
\bibfield{author}{\bibinfo{person}{Brett Daniel}, \bibinfo{person}{Tihomir
  Gvero}, {and} \bibinfo{person}{Darko Marinov}.}
  \bibinfo{year}{2010}\natexlab{}.
\newblock \showarticletitle{On test repair using symbolic execution}. In
  \bibinfo{booktitle}{{\em Proceedings of the 19th international symposium on
  Software testing and analysis}}. ACM, \bibinfo{pages}{207--218}.
\newblock


\bibitem[\protect\citeauthoryear{Dannenberg and Ernst}{Dannenberg and
  Ernst}{1982}]%
        {Dannenberg}
\bibfield{author}{\bibinfo{person}{R.B. Dannenberg} {and} \bibinfo{person}{G.W.
  Ernst}.} \bibinfo{year}{1982}\natexlab{}.
\newblock \showarticletitle{Formal Program Verification using Symbolic
  Execution}.
\newblock \bibinfo{journal}{{\em {IEEE} Transactions on Software
  Engineering\/}}  \bibinfo{volume}{8} (\bibinfo{year}{1982}).
\newblock
Issue 1.


\bibitem[\protect\citeauthoryear{Davis, Logemann, and Loveland}{Davis
  et~al\mbox{.}}{1962}]%
        {davis1962machine}
\bibfield{author}{\bibinfo{person}{Martin Davis}, \bibinfo{person}{George
  Logemann}, {and} \bibinfo{person}{Donald Loveland}.}
  \bibinfo{year}{1962}\natexlab{}.
\newblock \showarticletitle{A machine program for theorem-proving}.
\newblock \bibinfo{journal}{{\it Commun. ACM}} \bibinfo{volume}{5},
  \bibinfo{number}{7} (\bibinfo{year}{1962}), \bibinfo{pages}{394--397}.
\newblock


\bibitem[\protect\citeauthoryear{de~Moura and Bj{\o}rner}{de~Moura and
  Bj{\o}rner}{2008}]%
        {z3}
\bibfield{author}{\bibinfo{person}{Leonardo de Moura} {and}
  \bibinfo{person}{Nikolaj Bj{\o}rner}.} \bibinfo{year}{2008}\natexlab{}.
\newblock \showarticletitle{Z3: An Efficient SMT Solver}. In
  \bibinfo{booktitle}{{\em Tools and Algorithms for the Construction and
  Analysis of Systems}}, \bibfield{editor}{\bibinfo{person}{C.~R. Ramakrishnan}
  {and} \bibinfo{person}{Jakob Rehof}} (Eds.). \bibinfo{publisher}{Springer
  Berlin Heidelberg}, \bibinfo{address}{Berlin, Heidelberg},
  \bibinfo{pages}{337--340}.
\newblock
\showISBNx{978-3-540-78800-3}


\bibitem[\protect\citeauthoryear{Duchi, Hazan, and Singer}{Duchi
  et~al\mbox{.}}{2011}]%
        {duchi2011adaptive}
\bibfield{author}{\bibinfo{person}{John Duchi}, \bibinfo{person}{Elad Hazan},
  {and} \bibinfo{person}{Yoram Singer}.} \bibinfo{year}{2011}\natexlab{}.
\newblock \showarticletitle{Adaptive subgradient methods for online learning
  and stochastic optimization}.
\newblock \bibinfo{journal}{{\em Journal of Machine Learning Research\/}}
  \bibinfo{volume}{12}, \bibinfo{number}{Jul} (\bibinfo{year}{2011}),
  \bibinfo{pages}{2121--2159}.
\newblock


\bibitem[\protect\citeauthoryear{Ernst, Perkins, Guo, McCamant, Pacheco,
  Tschantz, and Xiao}{Ernst et~al\mbox{.}}{2007}]%
        {Daikon}
\bibfield{author}{\bibinfo{person}{Michael~D Ernst}, \bibinfo{person}{Jeff~H
  Perkins}, \bibinfo{person}{Philip~J Guo}, \bibinfo{person}{Stephen McCamant},
  \bibinfo{person}{Carlos Pacheco}, \bibinfo{person}{Matthew~S Tschantz}, {and}
  \bibinfo{person}{Chen Xiao}.} \bibinfo{year}{2007}\natexlab{}.
\newblock \showarticletitle{The Daikon system for dynamic detection of likely
  invariants}.
\newblock \bibinfo{journal}{{\em Science of Computer Programming\/}}
  \bibinfo{volume}{69}, \bibinfo{number}{1-3}, \bibinfo{pages}{35--45}.
\newblock


\bibitem[\protect\citeauthoryear{Funahashi}{Funahashi}{1989}]%
        {funahashi1989approximate}
\bibfield{author}{\bibinfo{person}{Ken-Ichi Funahashi}.}
  \bibinfo{year}{1989}\natexlab{}.
\newblock \showarticletitle{On the approximate realization of continuous
  mappings by neural networks}.
\newblock \bibinfo{journal}{{\em Neural networks\/}} \bibinfo{volume}{2},
  \bibinfo{number}{3} (\bibinfo{year}{1989}), \bibinfo{pages}{183--192}.
\newblock


\bibitem[\protect\citeauthoryear{Ganesh, Kie{\.z}un, Artzi, Guo, Hooimeijer,
  and Ernst}{Ganesh et~al\mbox{.}}{2011}]%
        {ganesh2011hampi}
\bibfield{author}{\bibinfo{person}{Vijay Ganesh}, \bibinfo{person}{Adam
  Kie{\.z}un}, \bibinfo{person}{Shay Artzi}, \bibinfo{person}{Philip~J Guo},
  \bibinfo{person}{Pieter Hooimeijer}, {and} \bibinfo{person}{Michael Ernst}.}
  \bibinfo{year}{2011}\natexlab{}.
\newblock \showarticletitle{HAMPI: A string solver for testing, analysis and
  vulnerability detection}. In \bibinfo{booktitle}{{\em International
  Conference on Computer Aided Verification}}. Springer,
  \bibinfo{pages}{1--19}.
\newblock


\bibitem[\protect\citeauthoryear{Garg, L{\"o}ding, Madhusudan, and Neider}{Garg
  et~al\mbox{.}}{2014}]%
        {garg2014ice}
\bibfield{author}{\bibinfo{person}{Pranav Garg}, \bibinfo{person}{Christof
  L{\"o}ding}, \bibinfo{person}{P Madhusudan}, {and} \bibinfo{person}{Daniel
  Neider}.} \bibinfo{year}{2014}\natexlab{}.
\newblock \showarticletitle{ICE: A robust framework for learning invariants}.
  In \bibinfo{booktitle}{{\em International Conference on Computer Aided
  Verification}}. Springer, \bibinfo{pages}{69--87}.
\newblock


\bibitem[\protect\citeauthoryear{Glorot, Bordes, and Bengio}{Glorot
  et~al\mbox{.}}{2011}]%
        {glorot2011deep}
\bibfield{author}{\bibinfo{person}{Xavier Glorot}, \bibinfo{person}{Antoine
  Bordes}, {and} \bibinfo{person}{Yoshua Bengio}.}
  \bibinfo{year}{2011}\natexlab{}.
\newblock \showarticletitle{Deep sparse rectifier neural networks}. In
  \bibinfo{booktitle}{{\em Proceedings of the Fourteenth International
  Conference on Artificial Intelligence and Statistics}}.
  \bibinfo{pages}{315--323}.
\newblock


\bibitem[\protect\citeauthoryear{Godefroid}{Godefroid}{2007}]%
        {godefroid}
\bibfield{author}{\bibinfo{person}{Patrice Godefroid}.}
  \bibinfo{year}{2007}\natexlab{}.
\newblock \showarticletitle{Compositional Dynamic Test Generation}. In
  \bibinfo{booktitle}{{\em Proceedings of 34th Symposium on Principles of
  Programming Languages ({POPL})}}.
\newblock


\bibitem[\protect\citeauthoryear{Godefroid, Klarlund, and Sen}{Godefroid
  et~al\mbox{.}}{2005}]%
        {DART}
\bibfield{author}{\bibinfo{person}{Patrice Godefroid}, \bibinfo{person}{Nils
  Klarlund}, {and} \bibinfo{person}{Koushik Sen}.}
  \bibinfo{year}{2005}\natexlab{}.
\newblock \showarticletitle{DART: Directed Automated Random Testing}. In
  \bibinfo{booktitle}{{\em Proceedings of International Symposium on
  Programming Language Design and Implementation ({PLDI})}}.
\newblock


\bibitem[\protect\citeauthoryear{Godefroid, Levin, and Molnar}{Godefroid
  et~al\mbox{.}}{2012}]%
        {godefroid2012sage}
\bibfield{author}{\bibinfo{person}{Patrice Godefroid},
  \bibinfo{person}{Michael~Y Levin}, {and} \bibinfo{person}{David Molnar}.}
  \bibinfo{year}{2012}\natexlab{}.
\newblock \showarticletitle{SAGE: whitebox fuzzing for security testing}.
\newblock \bibinfo{journal}{{\it Commun. ACM}} \bibinfo{volume}{55},
  \bibinfo{number}{3} (\bibinfo{year}{2012}), \bibinfo{pages}{40--44}.
\newblock


\bibitem[\protect\citeauthoryear{Godefroid, Levin, Molnar,
  et~al\mbox{.}}{Godefroid et~al\mbox{.}}{2008}]%
        {godefroid2008automated}
\bibfield{author}{\bibinfo{person}{Patrice Godefroid},
  \bibinfo{person}{Michael~Y Levin}, \bibinfo{person}{David~A Molnar},
  {et~al\mbox{.}}} \bibinfo{year}{2008}\natexlab{}.
\newblock \showarticletitle{Automated whitebox fuzz testing.}. In
  \bibinfo{booktitle}{{\em NDSS}}, Vol.~\bibinfo{volume}{8}.
  \bibinfo{pages}{151--166}.
\newblock


\bibitem[\protect\citeauthoryear{Godfrey and Gashler}{Godfrey and
  Gashler}{2015}]%
        {godfrey2015continuum}
\bibfield{author}{\bibinfo{person}{Luke~B Godfrey} {and}
  \bibinfo{person}{Michael~S Gashler}.} \bibinfo{year}{2015}\natexlab{}.
\newblock \showarticletitle{A continuum among logarithmic, linear, and
  exponential functions, and its potential to improve generalization in neural
  networks}. In \bibinfo{booktitle}{{\em Knowledge Discovery, Knowledge
  Engineering and Knowledge Management (IC3K), 2015 7th International Joint
  Conference on}}, Vol.~\bibinfo{volume}{1}. IEEE, \bibinfo{pages}{481--486}.
\newblock


\bibitem[\protect\citeauthoryear{Goodfellow, Shlens, and Szegedy}{Goodfellow
  et~al\mbox{.}}{2015}]%
        {goodfellow2014explaining}
\bibfield{author}{\bibinfo{person}{Ian~J Goodfellow}, \bibinfo{person}{Jonathon
  Shlens}, {and} \bibinfo{person}{Christian Szegedy}.}
  \bibinfo{year}{2015}\natexlab{}.
\newblock \showarticletitle{Explaining and harnessing adversarial examples}. In
  \bibinfo{booktitle}{{\em International Conference on Learning
  Representations}}.
\newblock


\bibitem[\protect\citeauthoryear{Gupta and Rybalchenko}{Gupta and
  Rybalchenko}{2009}]%
        {gupta2009invgen}
\bibfield{author}{\bibinfo{person}{Ashutosh Gupta} {and}
  \bibinfo{person}{Andrey Rybalchenko}.} \bibinfo{year}{2009}\natexlab{}.
\newblock \showarticletitle{Invgen: An efficient invariant generator}. In
  \bibinfo{booktitle}{{\em International Conference on Computer Aided
  Verification}}. Springer, \bibinfo{pages}{634--640}.
\newblock


\bibitem[\protect\citeauthoryear{Hornik}{Hornik}{1991}]%
        {hornik1991approximation}
\bibfield{author}{\bibinfo{person}{Kurt Hornik}.}
  \bibinfo{year}{1991}\natexlab{}.
\newblock \showarticletitle{Approximation capabilities of multilayer
  feedforward networks}.
\newblock \bibinfo{journal}{{\em Neural networks\/}} \bibinfo{volume}{4},
  \bibinfo{number}{2} (\bibinfo{year}{1991}), \bibinfo{pages}{251--257}.
\newblock


\bibitem[\protect\citeauthoryear{Jaffar, Murali, Navas, and Santosa}{Jaffar
  et~al\mbox{.}}{2012}]%
        {jaffar2012tracer}
\bibfield{author}{\bibinfo{person}{Joxan Jaffar},
  \bibinfo{person}{Vijayaraghavan Murali}, \bibinfo{person}{Jorge~A Navas},
  {and} \bibinfo{person}{Andrew~E Santosa}.} \bibinfo{year}{2012}\natexlab{}.
\newblock \showarticletitle{TRACER: A symbolic execution tool for
  verification}. In \bibinfo{booktitle}{{\em International Conference on
  Computer Aided Verification}}. Springer, \bibinfo{pages}{758--766}.
\newblock


\bibitem[\protect\citeauthoryear{Jeon, Micinski, and Foster}{Jeon
  et~al\mbox{.}}{2012}]%
        {jeon2012symdroid}
\bibfield{author}{\bibinfo{person}{Jinseong Jeon},
  \bibinfo{person}{Kristopher~K Micinski}, {and} \bibinfo{person}{Jeffrey~S
  Foster}.} \bibinfo{year}{2012}\natexlab{}.
\newblock \bibinfo{booktitle}{{\em SymDroid: Symbolic execution for Dalvik
  bytecode}}.
\newblock \bibinfo{type}{{T}echnical {R}eport}.
\newblock


\bibitem[\protect\citeauthoryear{Jhala and McMillan}{Jhala and
  McMillan}{2006}]%
        {jhala2006practical}
\bibfield{author}{\bibinfo{person}{Ranjit Jhala} {and}
  \bibinfo{person}{Kenneth~L McMillan}.} \bibinfo{year}{2006}\natexlab{}.
\newblock \showarticletitle{A practical and complete approach to predicate
  refinement}. In \bibinfo{booktitle}{{\em International Conference on Tools
  and Algorithms for the Construction and Analysis of Systems}}. Springer,
  \bibinfo{pages}{459--473}.
\newblock


\bibitem[\protect\citeauthoryear{King}{King}{1976a}]%
        {King}
\bibfield{author}{\bibinfo{person}{J.C. King}.}
  \bibinfo{year}{1976}\natexlab{a}.
\newblock \showarticletitle{Symbolic Execution and Program Testing}.
\newblock \bibinfo{journal}{{\it Commun. {ACM}}}  \bibinfo{volume}{19}
  (\bibinfo{year}{1976}).
\newblock
Issue 7.


\bibitem[\protect\citeauthoryear{King}{King}{1976b}]%
        {king1976symbolic}
\bibfield{author}{\bibinfo{person}{James~C King}.}
  \bibinfo{year}{1976}\natexlab{b}.
\newblock \showarticletitle{Symbolic execution and program testing}.
\newblock \bibinfo{journal}{{\it Commun. ACM}} \bibinfo{volume}{19},
  \bibinfo{number}{7} (\bibinfo{year}{1976}), \bibinfo{pages}{385--394}.
\newblock


\bibitem[\protect\citeauthoryear{Kingma and Ba}{Kingma and Ba}{2015}]%
        {kingma2014adam}
\bibfield{author}{\bibinfo{person}{Diederik~P Kingma} {and}
  \bibinfo{person}{Jimmy Ba}.} \bibinfo{year}{2015}\natexlab{}.
\newblock \showarticletitle{Adam: A method for stochastic optimization}. In
  \bibinfo{booktitle}{{\em International Conference on Learning
  Representations}}.
\newblock


\bibitem[\protect\citeauthoryear{Krishna, Puhrsch, and Wies}{Krishna
  et~al\mbox{.}}{2015}]%
        {krishna2015learning}
\bibfield{author}{\bibinfo{person}{Siddharth Krishna},
  \bibinfo{person}{Christian Puhrsch}, {and} \bibinfo{person}{Thomas Wies}.}
  \bibinfo{year}{2015}\natexlab{}.
\newblock \showarticletitle{Learning invariants using decision trees}.
\newblock \bibinfo{journal}{{\em arXiv preprint arXiv:1501.04725\/}}
  (\bibinfo{year}{2015}).
\newblock


\bibitem[\protect\citeauthoryear{Krizhevsky, Sutskever, and Hinton}{Krizhevsky
  et~al\mbox{.}}{2012}]%
        {krizhevsky2012imagenet}
\bibfield{author}{\bibinfo{person}{Alex Krizhevsky}, \bibinfo{person}{Ilya
  Sutskever}, {and} \bibinfo{person}{Geoffrey~E Hinton}.}
  \bibinfo{year}{2012}\natexlab{}.
\newblock \showarticletitle{Imagenet classification with deep convolutional
  neural networks}. In \bibinfo{booktitle}{{\em Advances in neural information
  processing systems}}. \bibinfo{pages}{1097--1105}.
\newblock


\bibitem[\protect\citeauthoryear{Kuznetsov, Kinder, Bucur, and
  Candea}{Kuznetsov et~al\mbox{.}}{2012}]%
        {DSM-kuznetsov}
\bibfield{author}{\bibinfo{person}{V. Kuznetsov}, \bibinfo{person}{J. Kinder},
  \bibinfo{person}{S. Bucur}, {and} \bibinfo{person}{G. Candea}.}
  \bibinfo{year}{2012}\natexlab{}.
\newblock \showarticletitle{Efficient state merging in symbolic execution}. In
  \bibinfo{booktitle}{{\em Proceedings of the 33rd ACM SIGPLAN Conference on
  Programming Language Design and Implementation ({PLDI})}}.
\newblock


\bibitem[\protect\citeauthoryear{Lawrence, Giles, Tsoi, and Back}{Lawrence
  et~al\mbox{.}}{1997}]%
        {lawrence1997face}
\bibfield{author}{\bibinfo{person}{Steve Lawrence}, \bibinfo{person}{C~Lee
  Giles}, \bibinfo{person}{Ah~Chung Tsoi}, {and} \bibinfo{person}{Andrew~D
  Back}.} \bibinfo{year}{1997}\natexlab{}.
\newblock \showarticletitle{Face recognition: A convolutional neural-network
  approach}.
\newblock \bibinfo{journal}{{\em IEEE transactions on neural networks\/}}
  \bibinfo{volume}{8}, \bibinfo{number}{1} (\bibinfo{year}{1997}),
  \bibinfo{pages}{98--113}.
\newblock


\bibitem[\protect\citeauthoryear{Li, Andreasen, and Ghosh}{Li
  et~al\mbox{.}}{2014}]%
        {li2014symjs}
\bibfield{author}{\bibinfo{person}{Guodong Li}, \bibinfo{person}{Esben
  Andreasen}, {and} \bibinfo{person}{Indradeep Ghosh}.}
  \bibinfo{year}{2014}\natexlab{}.
\newblock \showarticletitle{SymJS: automatic symbolic testing of JavaScript web
  applications}. In \bibinfo{booktitle}{{\em Proceedings of the 22nd ACM
  SIGSOFT International Symposium on Foundations of Software Engineering}}.
  ACM, \bibinfo{pages}{449--459}.
\newblock


\bibitem[\protect\citeauthoryear{Li, Liang, Qian, Hu, Bu, Yu, Chen, and Li}{Li
  et~al\mbox{.}}{2016}]%
        {li2016symbolic}
\bibfield{author}{\bibinfo{person}{Xin Li}, \bibinfo{person}{Yongjuan Liang},
  \bibinfo{person}{Hong Qian}, \bibinfo{person}{Yi-Qi Hu}, \bibinfo{person}{Lei
  Bu}, \bibinfo{person}{Yang Yu}, \bibinfo{person}{Xin Chen}, {and}
  \bibinfo{person}{Xuandong Li}.} \bibinfo{year}{2016}\natexlab{}.
\newblock \showarticletitle{Symbolic execution of complex program driven by
  machine learning based constraint solving}. In \bibinfo{booktitle}{{\em
  Proceedings of the 31st IEEE/ACM International Conference on Automated
  Software Engineering}}. ACM, \bibinfo{pages}{554--559}.
\newblock


\bibitem[\protect\citeauthoryear{Liang, Ganesh, Poupart, and Czarnecki}{Liang
  et~al\mbox{.}}{2016}]%
        {liang2016exponential}
\bibfield{author}{\bibinfo{person}{Jia~Hui Liang}, \bibinfo{person}{Vijay
  Ganesh}, \bibinfo{person}{Pascal Poupart}, {and} \bibinfo{person}{Krzysztof
  Czarnecki}.} \bibinfo{year}{2016}\natexlab{}.
\newblock \showarticletitle{Exponential Recency Weighted Average Branching
  Heuristic for SAT Solvers.}. In \bibinfo{booktitle}{{\em AAAI}}.
  \bibinfo{pages}{3434--3440}.
\newblock


\bibitem[\protect\citeauthoryear{Maas, Hannun, and Ng}{Maas
  et~al\mbox{.}}{2013}]%
        {maas2013rectifier}
\bibfield{author}{\bibinfo{person}{Andrew~L Maas}, \bibinfo{person}{Awni~Y
  Hannun}, {and} \bibinfo{person}{Andrew~Y Ng}.}
  \bibinfo{year}{2013}\natexlab{}.
\newblock \showarticletitle{Rectifier nonlinearities improve neural network
  acoustic models}. In \bibinfo{booktitle}{{\em Proc. icml}},
  Vol.~\bibinfo{volume}{30}. \bibinfo{pages}{3}.
\newblock


\bibitem[\protect\citeauthoryear{McMillan}{McMillan}{2003}]%
        {mcmillan2013interpolation}
\bibfield{author}{\bibinfo{person}{Ken McMillan}.}
  \bibinfo{year}{2003}\natexlab{}.
\newblock \showarticletitle{Interpolation and SAT-based Model Checking}. In
  \bibinfo{booktitle}{{\em International Conference on Computer Aided
  Verification}}.
\newblock


\bibitem[\protect\citeauthoryear{Medsker and Jain}{Medsker and Jain}{2001}]%
        {medsker2001recurrent}
\bibfield{author}{\bibinfo{person}{LR Medsker} {and} \bibinfo{person}{LC
  Jain}.} \bibinfo{year}{2001}\natexlab{}.
\newblock \showarticletitle{Recurrent neural networks}.
\newblock \bibinfo{journal}{{\em Design and Applications\/}}
  \bibinfo{volume}{5} (\bibinfo{year}{2001}).
\newblock


\bibitem[\protect\citeauthoryear{Mikolov, Karafi{\'a}t, Burget,
  {\v{C}}ernock{\`y}, and Khudanpur}{Mikolov et~al\mbox{.}}{2010}]%
        {mikolov2010recurrent}
\bibfield{author}{\bibinfo{person}{Tom{\'a}{\v{s}} Mikolov},
  \bibinfo{person}{Martin Karafi{\'a}t}, \bibinfo{person}{Luk{\'a}{\v{s}}
  Burget}, \bibinfo{person}{Jan {\v{C}}ernock{\`y}}, {and}
  \bibinfo{person}{Sanjeev Khudanpur}.} \bibinfo{year}{2010}\natexlab{}.
\newblock \showarticletitle{Recurrent neural network based language model}. In
  \bibinfo{booktitle}{{\em Eleventh Annual Conference of the International
  Speech Communication Association}}.
\newblock


\bibitem[\protect\citeauthoryear{Min{\'e}}{Min{\'e}}{2004}]%
        {mine2004weakly}
\bibfield{author}{\bibinfo{person}{Antoine Min{\'e}}.}
  \bibinfo{year}{2004}\natexlab{}.
\newblock {\em \bibinfo{title}{Weakly relational numerical abstract domains}}.
\newblock \bibinfo{thesistype}{Ph.D. Dissertation}. \bibinfo{school}{Ecole
  Polytechnique X}.
\newblock


\bibitem[\protect\citeauthoryear{Moskewicz, Madigan, Zhao, Zhang, and
  Malik}{Moskewicz et~al\mbox{.}}{2001}]%
        {moskewicz2001chaff}
\bibfield{author}{\bibinfo{person}{Matthew~W Moskewicz},
  \bibinfo{person}{Conor~F Madigan}, \bibinfo{person}{Ying Zhao},
  \bibinfo{person}{Lintao Zhang}, {and} \bibinfo{person}{Sharad Malik}.}
  \bibinfo{year}{2001}\natexlab{}.
\newblock \showarticletitle{Chaff: Engineering an efficient SAT solver}. In
  \bibinfo{booktitle}{{\em Proceedings of the 38th annual Design Automation
  Conference}}. ACM, \bibinfo{pages}{530--535}.
\newblock


\bibitem[\protect\citeauthoryear{Narodytska, Kasiviswanathan, Ryzhyk, Sagiv,
  and Walsh}{Narodytska et~al\mbox{.}}{2017}]%
        {narodytska2017verifying}
\bibfield{author}{\bibinfo{person}{Nina Narodytska},
  \bibinfo{person}{Shiva~Prasad Kasiviswanathan}, \bibinfo{person}{Leonid
  Ryzhyk}, \bibinfo{person}{Mooly Sagiv}, {and} \bibinfo{person}{Toby Walsh}.}
  \bibinfo{year}{2017}\natexlab{}.
\newblock \showarticletitle{Verifying properties of binarized deep neural
  networks}.
\newblock \bibinfo{journal}{{\em arXiv preprint arXiv:1709.06662\/}}
  (\bibinfo{year}{2017}).
\newblock


\bibitem[\protect\citeauthoryear{Nesterov and Polyak}{Nesterov and
  Polyak}{2006}]%
        {nesterov2006cubic}
\bibfield{author}{\bibinfo{person}{Yurii Nesterov} {and}
  \bibinfo{person}{Boris~T Polyak}.} \bibinfo{year}{2006}\natexlab{}.
\newblock \showarticletitle{Cubic regularization of Newton method and its
  global performance}.
\newblock \bibinfo{journal}{{\em Mathematical Programming\/}}
  \bibinfo{volume}{108}, \bibinfo{number}{1} (\bibinfo{year}{2006}),
  \bibinfo{pages}{177--205}.
\newblock


\bibitem[\protect\citeauthoryear{Nguyen, Qi, Roychoudhury, and Chandra}{Nguyen
  et~al\mbox{.}}{2013}]%
        {nguyen2013semfix}
\bibfield{author}{\bibinfo{person}{Hoang Duong~Thien Nguyen},
  \bibinfo{person}{Dawei Qi}, \bibinfo{person}{Abhik Roychoudhury}, {and}
  \bibinfo{person}{Satish Chandra}.} \bibinfo{year}{2013}\natexlab{}.
\newblock \showarticletitle{Semfix: Program repair via semantic analysis}. In
  \bibinfo{booktitle}{{\em Proceedings of the 2013 International Conference on
  Software Engineering}}. IEEE Press, \bibinfo{pages}{772--781}.
\newblock


\bibitem[\protect\citeauthoryear{Nguyen, Antonopoulos, Ruef, and Hicks}{Nguyen
  et~al\mbox{.}}{2017}]%
        {DIG2}
\bibfield{author}{\bibinfo{person}{ThanhVu Nguyen}, \bibinfo{person}{Timos
  Antonopoulos}, \bibinfo{person}{Andrew Ruef}, {and} \bibinfo{person}{Michael
  Hicks}.} \bibinfo{year}{2017}\natexlab{}.
\newblock \showarticletitle{Counterexample-guided approach to finding numerical
  invariants}. In \bibinfo{booktitle}{{\em Proceedings of the 2017 11th Joint
  Meeting on Foundations of Software Engineering}}. ACM,
  \bibinfo{pages}{605--615}.
\newblock


\bibitem[\protect\citeauthoryear{Nguyen, Kapur, Weimer, and Forrest}{Nguyen
  et~al\mbox{.}}{2014}]%
        {DIG}
\bibfield{author}{\bibinfo{person}{Thanhvu Nguyen}, \bibinfo{person}{Deepak
  Kapur}, \bibinfo{person}{Westley Weimer}, {and} \bibinfo{person}{Stephanie
  Forrest}.} \bibinfo{year}{2014}\natexlab{}.
\newblock \showarticletitle{DIG: a dynamic invariant generator for polynomial
  and array invariants}.
\newblock \bibinfo{journal}{{\em ACM Transactions on Software Engineering and
  Methodology (TOSEM)\/}} \bibinfo{volume}{23}, \bibinfo{number}{4},
  \bibinfo{pages}{30}.
\newblock


\bibitem[\protect\citeauthoryear{Padhi and Millstein}{Padhi and
  Millstein}{2017}]%
        {padhi2017data}
\bibfield{author}{\bibinfo{person}{Saswat Padhi} {and} \bibinfo{person}{Todd
  Millstein}.} \bibinfo{year}{2017}\natexlab{}.
\newblock \showarticletitle{Data-Driven Loop Invariant Inference with Automatic
  Feature Synthesis}.
\newblock \bibinfo{journal}{{\em arXiv preprint arXiv:1707.02029\/}}
  (\bibinfo{year}{2017}).
\newblock


\bibitem[\protect\citeauthoryear{Papernot, McDaniel, Jha, Fredrikson, Celik,
  and Swami}{Papernot et~al\mbox{.}}{2016}]%
        {papernot2016limitations}
\bibfield{author}{\bibinfo{person}{Nicolas Papernot}, \bibinfo{person}{Patrick
  McDaniel}, \bibinfo{person}{Somesh Jha}, \bibinfo{person}{Matt Fredrikson},
  \bibinfo{person}{Z~Berkay Celik}, {and} \bibinfo{person}{Ananthram Swami}.}
  \bibinfo{year}{2016}\natexlab{}.
\newblock \showarticletitle{The limitations of deep learning in adversarial
  settings}. In \bibinfo{booktitle}{{\em Security and Privacy (EuroS\&P), 2016
  IEEE European Symposium on}}. IEEE, \bibinfo{pages}{372--387}.
\newblock


\bibitem[\protect\citeauthoryear{Perkins, Kim, Larsen, Amarasinghe, Bachrach,
  Carbin, Pacheco, Sherwood, Sidiroglou, Sullivan, et~al\mbox{.}}{Perkins
  et~al\mbox{.}}{2009}]%
        {perkins2009automatically}
\bibfield{author}{\bibinfo{person}{Jeff~H Perkins}, \bibinfo{person}{Sunghun
  Kim}, \bibinfo{person}{Sam Larsen}, \bibinfo{person}{Saman Amarasinghe},
  \bibinfo{person}{Jonathan Bachrach}, \bibinfo{person}{Michael Carbin},
  \bibinfo{person}{Carlos Pacheco}, \bibinfo{person}{Frank Sherwood},
  \bibinfo{person}{Stelios Sidiroglou}, \bibinfo{person}{Greg Sullivan},
  {et~al\mbox{.}}} \bibinfo{year}{2009}\natexlab{}.
\newblock \showarticletitle{Automatically patching errors in deployed
  software}. In \bibinfo{booktitle}{{\em Proceedings of the ACM SIGOPS 22nd
  symposium on Operating systems principles}}. ACM, \bibinfo{pages}{87--102}.
\newblock


\bibitem[\protect\citeauthoryear{Qi, Nguyen, and Roychoudhury}{Qi
  et~al\mbox{.}}{2013}]%
        {PESO}
\bibfield{author}{\bibinfo{person}{D. Qi}, \bibinfo{person}{H.D.T Nguyen},
  {and} \bibinfo{person}{A. Roychoudhury}.} \bibinfo{year}{2013}\natexlab{}.
\newblock \showarticletitle{Path Exploration using Symbolic Output}.
\newblock \bibinfo{journal}{{\em ACM Transactions on Software Engineering and
  Methodology ({TOSEM})\/}}  \bibinfo{volume}{22} (\bibinfo{year}{2013}).
\newblock
Issue 4.


\bibitem[\protect\citeauthoryear{Qian}{Qian}{1999}]%
        {qian1999momentum}
\bibfield{author}{\bibinfo{person}{Ning Qian}.}
  \bibinfo{year}{1999}\natexlab{}.
\newblock \showarticletitle{On the momentum term in gradient descent learning
  algorithms}.
\newblock \bibinfo{journal}{{\em Neural networks\/}} \bibinfo{volume}{12},
  \bibinfo{number}{1} (\bibinfo{year}{1999}), \bibinfo{pages}{145--151}.
\newblock


\bibitem[\protect\citeauthoryear{Rodr{\'\i}guez-Carbonell and
  Kapur}{Rodr{\'\i}guez-Carbonell and Kapur}{2007a}]%
        {rodriguez2007automatic}
\bibfield{author}{\bibinfo{person}{Enric Rodr{\'\i}guez-Carbonell} {and}
  \bibinfo{person}{Deepak Kapur}.} \bibinfo{year}{2007}\natexlab{a}.
\newblock \showarticletitle{Automatic generation of polynomial invariants of
  bounded degree using abstract interpretation}.
\newblock \bibinfo{journal}{{\em Science of Computer Programming\/}}
  \bibinfo{volume}{64}, \bibinfo{number}{1} (\bibinfo{year}{2007}),
  \bibinfo{pages}{54--75}.
\newblock


\bibitem[\protect\citeauthoryear{Rodr{\'\i}guez-Carbonell and
  Kapur}{Rodr{\'\i}guez-Carbonell and Kapur}{2007b}]%
        {rodriguez2007generating}
\bibfield{author}{\bibinfo{person}{Enric Rodr{\'\i}guez-Carbonell} {and}
  \bibinfo{person}{Deepak Kapur}.} \bibinfo{year}{2007}\natexlab{b}.
\newblock \showarticletitle{Generating all polynomial invariants in simple
  loops}.
\newblock \bibinfo{journal}{{\em Journal of Symbolic Computation\/}}
  \bibinfo{volume}{42}, \bibinfo{number}{4} (\bibinfo{year}{2007}),
  \bibinfo{pages}{443--476}.
\newblock


\bibitem[\protect\citeauthoryear{Ruder}{Ruder}{2016}]%
        {ruder2016overview}
\bibfield{author}{\bibinfo{person}{Sebastian Ruder}.}
  \bibinfo{year}{2016}\natexlab{}.
\newblock \showarticletitle{An overview of gradient descent optimization
  algorithms}.
\newblock \bibinfo{journal}{{\em CoRR, abs/1609.04747\/}}
  (\bibinfo{year}{2016}).
\newblock


\bibitem[\protect\citeauthoryear{Rumelhart, Hinton, and Williams}{Rumelhart
  et~al\mbox{.}}{1985}]%
        {rumelhart1985learning}
\bibfield{author}{\bibinfo{person}{David~E Rumelhart},
  \bibinfo{person}{Geoffrey~E Hinton}, {and} \bibinfo{person}{Ronald~J
  Williams}.} \bibinfo{year}{1985}\natexlab{}.
\newblock \bibinfo{booktitle}{{\em Learning internal representations by error
  propagation}}.
\newblock \bibinfo{type}{{T}echnical {R}eport}.
  \bibinfo{institution}{California Univ San Diego La Jolla Inst for Cognitive
  Science}.
\newblock


\bibitem[\protect\citeauthoryear{Sankaranarayanan, Chaudhuri,
  Ivan{\v{c}}i{\'c}, and Gupta}{Sankaranarayanan et~al\mbox{.}}{2008}]%
        {sankaranarayanan2008dynamic}
\bibfield{author}{\bibinfo{person}{Sriram Sankaranarayanan},
  \bibinfo{person}{Swarat Chaudhuri}, \bibinfo{person}{Franjo
  Ivan{\v{c}}i{\'c}}, {and} \bibinfo{person}{Aarti Gupta}.}
  \bibinfo{year}{2008}\natexlab{}.
\newblock \showarticletitle{Dynamic inference of likely data preconditions over
  predicates by tree learning}. In \bibinfo{booktitle}{{\em Proceedings of the
  2008 international symposium on Software testing and analysis}}. ACM,
  \bibinfo{pages}{295--306}.
\newblock


\bibitem[\protect\citeauthoryear{Saxena, Akhawe, Hanna, Mao, McCamant, and
  Song}{Saxena et~al\mbox{.}}{2010a}]%
        {saxena2010symbolic}
\bibfield{author}{\bibinfo{person}{Prateek Saxena}, \bibinfo{person}{Devdatta
  Akhawe}, \bibinfo{person}{Steve Hanna}, \bibinfo{person}{Feng Mao},
  \bibinfo{person}{Stephen McCamant}, {and} \bibinfo{person}{Dawn Song}.}
  \bibinfo{year}{2010}\natexlab{a}.
\newblock \showarticletitle{A symbolic execution framework for javascript}. In
  \bibinfo{booktitle}{{\em Security and Privacy (SP), 2010 IEEE Symposium on}}.
  IEEE, \bibinfo{pages}{513--528}.
\newblock


\bibitem[\protect\citeauthoryear{Saxena, Akhawe, Hanna, Mao, McCamant, and
  Song}{Saxena et~al\mbox{.}}{2010b}]%
        {kudzu}
\bibfield{author}{\bibinfo{person}{Prateek Saxena}, \bibinfo{person}{Devdatta
  Akhawe}, \bibinfo{person}{Steve Hanna}, \bibinfo{person}{Feng Mao},
  \bibinfo{person}{Stephen McCamant}, {and} \bibinfo{person}{Dawn Song}.}
  \bibinfo{year}{2010}\natexlab{b}.
\newblock \showarticletitle{A Symbolic Execution Framework for JavaScript}. In
  \bibinfo{booktitle}{{\em Proceedings of the 2010 IEEE Symposium on Security
  and Privacy}} {\em (\bibinfo{series}{SP '10})}. \bibinfo{publisher}{IEEE
  Computer Society}, \bibinfo{address}{Washington, DC, USA},
  \bibinfo{pages}{513--528}.
\newblock
\showISBNx{978-0-7695-4035-1}
\showDOI{%
\url{https://doi.org/10.1109/SP.2010.38}}


\bibitem[\protect\citeauthoryear{Saxena, Poosankam, McCamant, and Song}{Saxena
  et~al\mbox{.}}{2009}]%
        {LESE}
\bibfield{author}{\bibinfo{person}{Prateek Saxena}, \bibinfo{person}{Pongsin
  Poosankam}, \bibinfo{person}{Stephen McCamant}, {and} \bibinfo{person}{Dawn
  Song}.} \bibinfo{year}{2009}\natexlab{}.
\newblock \showarticletitle{Loop-extended symbolic execution on binary
  programs}. In \bibinfo{booktitle}{{\em Proceedings of the eighteenth
  international symposium on Software testing and analysis}}. ACM,
  \bibinfo{pages}{225--236}.
\newblock


\bibitem[\protect\citeauthoryear{Sen, Necula, Gong, and Choi}{Sen
  et~al\mbox{.}}{2015}]%
        {multiSE}
\bibfield{author}{\bibinfo{person}{K. Sen}, \bibinfo{person}{G. Necula},
  \bibinfo{person}{L. Gong}, {and} \bibinfo{person}{W. Choi}.}
  \bibinfo{year}{2015}\natexlab{}.
\newblock \showarticletitle{multiSE: Multi-path Symbolic Execution}. In
  \bibinfo{booktitle}{{\em International Symposium on Foundations of Software
  Engineering}}.
\newblock


\bibitem[\protect\citeauthoryear{Siegel, Zheng, Luo, Zirkel, Marianiello,
  Edenhofner, Dwyer, and Rogers}{Siegel et~al\mbox{.}}{2015}]%
        {siegel2015civl}
\bibfield{author}{\bibinfo{person}{Stephen~F Siegel}, \bibinfo{person}{Manchun
  Zheng}, \bibinfo{person}{Ziqing Luo}, \bibinfo{person}{Timothy~K Zirkel},
  \bibinfo{person}{Andre~V Marianiello}, \bibinfo{person}{John~G Edenhofner},
  \bibinfo{person}{Matthew~B Dwyer}, {and} \bibinfo{person}{Michael~S Rogers}.}
  \bibinfo{year}{2015}\natexlab{}.
\newblock \showarticletitle{CIVL: the concurrency intermediate verification
  language}. In \bibinfo{booktitle}{{\em Proceedings of the International
  Conference for High Performance Computing, Networking, Storage and
  Analysis}}. ACM, \bibinfo{pages}{61}.
\newblock


\bibitem[\protect\citeauthoryear{Silva and Sakallah}{Silva and
  Sakallah}{1997}]%
        {silva1997grasp}
\bibfield{author}{\bibinfo{person}{Jo{\~a}o P~Marques Silva} {and}
  \bibinfo{person}{Karem~A Sakallah}.} \bibinfo{year}{1997}\natexlab{}.
\newblock \showarticletitle{GRASP—a new search algorithm for satisfiability}.
  In \bibinfo{booktitle}{{\em Proceedings of the 1996 IEEE/ACM international
  conference on Computer-aided design}}. IEEE Computer Society,
  \bibinfo{pages}{220--227}.
\newblock


\bibitem[\protect\citeauthoryear{Sorensen}{Sorensen}{1982}]%
        {sorensen1982newton}
\bibfield{author}{\bibinfo{person}{Danny~C Sorensen}.}
  \bibinfo{year}{1982}\natexlab{}.
\newblock \showarticletitle{Newton’s method with a model trust region
  modification}.
\newblock \bibinfo{journal}{{\it SIAM J. Numer. Anal.}} \bibinfo{volume}{19},
  \bibinfo{number}{2} (\bibinfo{year}{1982}), \bibinfo{pages}{409--426}.
\newblock


\bibitem[\protect\citeauthoryear{Srivastava, Hinton, Krizhevsky, Sutskever, and
  Salakhutdinov}{Srivastava et~al\mbox{.}}{2014}]%
        {srivastava2014dropout}
\bibfield{author}{\bibinfo{person}{Nitish Srivastava},
  \bibinfo{person}{Geoffrey Hinton}, \bibinfo{person}{Alex Krizhevsky},
  \bibinfo{person}{Ilya Sutskever}, {and} \bibinfo{person}{Ruslan
  Salakhutdinov}.} \bibinfo{year}{2014}\natexlab{}.
\newblock \showarticletitle{Dropout: A simple way to prevent neural networks
  from overfitting}.
\newblock \bibinfo{journal}{{\em The Journal of Machine Learning Research\/}}
  \bibinfo{volume}{15}, \bibinfo{number}{1} (\bibinfo{year}{2014}),
  \bibinfo{pages}{1929--1958}.
\newblock


\bibitem[\protect\citeauthoryear{Tikhonov}{Tikhonov}{1963}]%
        {tikhonov1963solution}
\bibfield{author}{\bibinfo{person}{Andrei~Nikolaevich Tikhonov}.}
  \bibinfo{year}{1963}\natexlab{}.
\newblock \showarticletitle{On the solution of ill-posed problems and the
  method of regularization}. In \bibinfo{booktitle}{{\em Doklady Akademii
  Nauk}}, Vol.~\bibinfo{volume}{151}. Russian Academy of Sciences,
  \bibinfo{pages}{501--504}.
\newblock


\bibitem[\protect\citeauthoryear{Xie, Tillmann, de~Halleux, and Schulte}{Xie
  et~al\mbox{.}}{2009}]%
        {xie2009fitness}
\bibfield{author}{\bibinfo{person}{Tao Xie}, \bibinfo{person}{Nikolai
  Tillmann}, \bibinfo{person}{Jonathan de Halleux}, {and}
  \bibinfo{person}{Wolfram Schulte}.} \bibinfo{year}{2009}\natexlab{}.
\newblock \showarticletitle{Fitness-guided path exploration in dynamic symbolic
  execution}. In \bibinfo{booktitle}{{\em Dependable Systems \& Networks, 2009.
  DSN'09. IEEE/IFIP International Conference on}}. IEEE,
  \bibinfo{pages}{359--368}.
\newblock


\bibitem[\protect\citeauthoryear{Xie, Chen, Liu, Le, and Li}{Xie
  et~al\mbox{.}}{2016}]%
        {xie2016proteus}
\bibfield{author}{\bibinfo{person}{Xiaofei Xie}, \bibinfo{person}{Bihuan Chen},
  \bibinfo{person}{Yang Liu}, \bibinfo{person}{Wei Le}, {and}
  \bibinfo{person}{Xiaohong Li}.} \bibinfo{year}{2016}\natexlab{}.
\newblock \showarticletitle{Proteus: computing disjunctive loop summary via
  path dependency analysis}. In \bibinfo{booktitle}{{\em Proceedings of the
  2016 24th ACM SIGSOFT International Symposium on Foundations of Software
  Engineering}}. ACM, \bibinfo{pages}{61--72}.
\newblock


\bibitem[\protect\citeauthoryear{Yao, Rosasco, and Caponnetto}{Yao
  et~al\mbox{.}}{2007}]%
        {yao2007early}
\bibfield{author}{\bibinfo{person}{Yuan Yao}, \bibinfo{person}{Lorenzo
  Rosasco}, {and} \bibinfo{person}{Andrea Caponnetto}.}
  \bibinfo{year}{2007}\natexlab{}.
\newblock \showarticletitle{On early stopping in gradient descent learning}.
\newblock \bibinfo{journal}{{\em Constructive Approximation\/}}
  \bibinfo{volume}{26}, \bibinfo{number}{2} (\bibinfo{year}{2007}),
  \bibinfo{pages}{289--315}.
\newblock


\bibitem[\protect\citeauthoryear{Yu, Qian, and Hu}{Yu et~al\mbox{.}}{2016}]%
        {yu2016derivative}
\bibfield{author}{\bibinfo{person}{Yang Yu}, \bibinfo{person}{Hong Qian}, {and}
  \bibinfo{person}{Yi-Qi Hu}.} \bibinfo{year}{2016}\natexlab{}.
\newblock \showarticletitle{Derivative-Free Optimization via Classification.}.
  In \bibinfo{booktitle}{{\em AAAI}}, Vol.~\bibinfo{volume}{16}.
  \bibinfo{pages}{2286--2292}.
\newblock


\bibitem[\protect\citeauthoryear{Zheng, Zhang, and Ganesh}{Zheng
  et~al\mbox{.}}{2013}]%
        {zheng2013z3}
\bibfield{author}{\bibinfo{person}{Yunhui Zheng}, \bibinfo{person}{Xiangyu
  Zhang}, {and} \bibinfo{person}{Vijay Ganesh}.}
  \bibinfo{year}{2013}\natexlab{}.
\newblock \showarticletitle{Z3-str: A z3-based string solver for web
  application analysis}. In \bibinfo{booktitle}{{\em Proceedings of the 2013
  9th Joint Meeting on Foundations of Software Engineering}}. ACM,
  \bibinfo{pages}{114--124}.
\newblock


\bibitem[\protect\citeauthoryear{Zitser, Lippmann, and Leek}{Zitser
  et~al\mbox{.}}{2004}]%
        {LESE_benchmark}
\bibfield{author}{\bibinfo{person}{Misha Zitser}, \bibinfo{person}{Richard
  Lippmann}, {and} \bibinfo{person}{Tim Leek}.}
  \bibinfo{year}{2004}\natexlab{}.
\newblock \showarticletitle{Testing static analysis tools using exploitable
  buffer overflows from open source code}. In \bibinfo{booktitle}{{\em ACM
  SIGSOFT Software Engineering Notes}}, Vol.~\bibinfo{volume}{29}. ACM,
  \bibinfo{pages}{97--106}.
\newblock


\end{thebibliography}
\appendix
\section{Appendix}
\label{sec:appx}

\subsection{Neural Constraint Analysis}
~\label{sec:analyze_nn}
We analyze the learned neural constraints by observing the trained weights and bias of neural network. Given a set of variables as the input to the neural network, if the input variable is not related with the output variable, the weight between the input and output variable is zero; otherwise, it is larger than zero. For example, the length of vulnerable buffer in program \texttt{Bind1} is controlled by \texttt{dlen} field which is the $43^{th}$ byte of of DNS queries, because the weight for this input variable is $0.99$ which has the largest absolute value compared with other fields.


\end{document}